# Parameterized Complexity in Multiple-Interval Graphs: Domination, Partition, Separation, Irredundancy[*]

Minghui Jiang[†]
Department of Computer Science, Utah State University,
Logan, UT 84322, USA
mjiang@cc.usu.edu

Yong Zhang
Department of Computer Science, Kutztown University of PA,
Kutztown, PA 19530, USA
zhang@kutztown.edu

October 23, 2018

**Abstract**

We show that the problem $k$-DOMINATING SET and its several variants including $k$-CONNECTED DOMINATING SET, $k$-INDEPENDENT DOMINATING SET, and $k$-DOMINATING CLIQUE, when parameterized by the solution size $k$, are W[1]-hard in either multiple-interval graphs or their complements or both. On the other hand, we show that these problems belong to W[1] when restricted to multiple-interval graphs and their complements. This answers an open question of Fellows et al. In sharp contrast, we show that $d$-DISTANCE $k$-DOMINATING SET for $d \geq 2$ is W[2]-complete in multiple-interval graphs and their complements. We also show that $k$-PERFECT CODE and $d$-DISTANCE $k$-PERFECT CODE for $d \geq 2$ are W[1]-complete even in unit 2-track interval graphs. In addition, we present various new results on the parameterized complexities of $k$-VERTEX CLIQUE PARTITION and $k$-SEPARATING VERTICES in multiple-interval graphs and their complements, and present a very simple alternative proof of the W[1]-hardness of $k$-IRREDUNDANT SET in general graphs.

## 1 Introduction

We introduce some basic definitions. The *intersection graph* $\Omega(\mathcal{F})$ of a family of sets $\mathcal{F} = \{S_1, \ldots, S_n\}$ is the graph with $\mathcal{F}$ as the vertex set and with two different vertices $S_i$ and $S_j$ adjacent if and only if $S_i \cap S_j \neq \emptyset$; the family $\mathcal{F}$ is called a *representation* of the graph $\Omega(\mathcal{F})$. Let $t \geq 2$ be an integer. A *$t$-interval graph* is the intersection graph of a family of $t$-intervals, where each *$t$-interval* is the union of $t$ disjoint intervals in the real line. A *$t$-track interval graph* is the intersection graph of a family of $t$-track intervals, where each *$t$-track interval* is the union of $t$ disjoint intervals on $t$ disjoint parallel lines called tracks, one interval on each track. Note that the $t$ disjoint tracks for a $t$-track interval graph can be viewed as $t$ disjoint "host intervals" in the real line for a $t$-interval graph. Thus $t$-track interval graphs are a subclass of $t$-interval graphs. If a $t$-interval graph has a representation in which all intervals have unit lengths, then the graph is a *unit $t$-interval graph*. If a $t$-interval graph has a representation in which the $t$ disjoint intervals of each $t$-interval have the same length (although the intervals from different $t$-intervals may have different lengths), then the graph is a *balanced $t$-interval graph.* Similarly we define unit $t$-track interval graphs and balanced $t$-track interval graphs. We refer to Figure 1 and Figure 2 for two examples.

---

[*]A preliminary version of this article appeared in two parts in COCOON 2011 [21] and IPEC 2011 [22].
[†]Supported in part by NSF grant DBI-0743670.

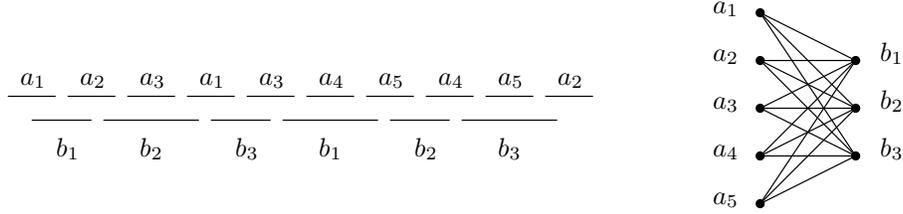

Figure 1: A 2-interval representation of the graph $K_{5,3}$.

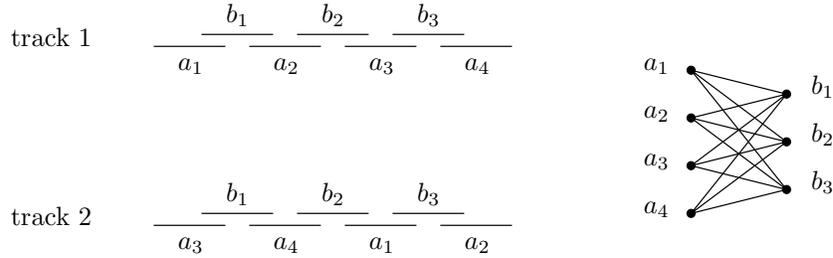

Figure 2: A unit 2-track interval representation of the graph $K_{4,3}$.

As generalizations of the ubiquitous interval graphs, multiple-interval graphs such as $t$-interval graphs and $t$-track interval graphs have numerous applications, traditionally to scheduling and resource allocation [26, 1], and more recently to bioinformatics [5, 18]. For this reason, a systematic study of various classical optimization problems in multiple-interval graphs has been undertaken by several groups of researchers. In terms of approximability, Bar-Yehuda et al. [1] presented a $2t$-approximation algorithm for MAXIMUM INDEPENDENT SET in $t$-interval graphs, and Butman et al. [2] presented approximation algorithms for MINIMUM VERTEX COVER, MINIMUM DOMINATING SET, and MAXIMUM CLIQUE in $t$-interval graphs with approximation ratios $2 - 1/t$, $t^2$, and $(t^2 - t + 1)/2$, respectively.

Fellows et al. [11] initiated the study of multiple-interval graph problems from the perspective of parameterized complexity. In general graphs, the four problems $k$-VERTEX COVER, $k$-INDEPENDENT SET, $k$-CLIQUE, and $k$-DOMINATING SET, parameterized by the solution size $k$, are exemplary problems in parameterized complexity theory [9]: it is well-known that $k$-VERTEX COVER is in FPT, $k$-INDEPENDENT SET and $k$-CLIQUE are W[1]-complete, and $k$-DOMINATING SET is W[2]-complete. Since $t$-interval graphs are a special class of graphs, all FPT algorithms for $k$-VERTEX COVER in general graphs immediately carry over to $t$-interval graphs. On the other hand, the parameterized complexities of $k$-INDEPENDENT SET, $k$-CLIQUE, and $k$-DOMINATING SET in $t$-interval graphs are not at all obvious. Indeed, in general graphs, $k$-INDEPENDENT SET and $k$-CLIQUE are essentially the same problem (the problem $k$-INDEPENDENT SET in any graph $G$ is the same as the problem $k$-CLIQUE in the complement graph $\overline{G}$), but in $t$-interval graphs, they manifest different parameterized complexities. Fellows et al. [11] showed that $k$-INDEPENDENT SET in $t$-interval graphs is W[1]-hard for any $t \geq 2$, then, in sharp contrast, gave an FPT algorithm for $k$-CLIQUE in $t$-interval graphs parameterized by both $k$ and $t$. Fellows et al. [11] also showed that $k$-DOMINATING SET in $t$-interval graphs is W[1]-hard for any $t \geq 2$. Recently, Jiang [19] strengthened the two hardness results for $t$-interval graphs, and showed that $k$-INDEPENDENT SET and $k$-DOMINATING SET remain W[1]-hard even in unit $t$-track interval graphs for any $t \geq 2$. In particular, we have the following theorem on the parameterized complexity of $k$-DOMINATING SET in unit 2-track interval graphs:

**Theorem 1** (Jiang 2010 [19])**.** $k$-DOMINATING SET *in unit 2-track interval graphs is W[1]-hard with parameter $k$.*

The lack of symmetry in the parameterized complexities of $k$-INDEPENDENT SET and $k$-CLIQUE in multiple-interval graphs and their complements leads to a natural question about $k$-DOMINATING SET, which is known to be W[1]-hard in multiple-interval graphs: Is it still W[1]-hard in the complements of multiple-interval graphs? Our following theorem (here "co-3-track interval graphs" denotes "complements of 3-track interval graphs") gives a positive answer:

**Theorem 2.** $k$-DOMINATING SET *in co-3-track interval graphs is W[1]-hard with parameter $k$.*

A *connected dominating set* in a graph $G$ is a dominating set $S$ in $G$ such that the induced subgraph $G(S)$ is connected. An *independent dominating set* in a graph $G$ is both a dominating set and an independent set in $G$. A *dominating clique* in a graph $G$ is both a dominating set and a clique in $G$. With connectivity taken in account, the problem $k$-DOMINATING SET has three important variants: $k$-CONNECTED DOMINATING SET, $k$-INDEPENDENT DOMINATING SET, and $k$-DOMINATING CLIQUE. Recall the sharp contrast in parameterized complexities of the two problems $k$-INDEPENDENT SET and $k$-CLIQUE in multiple-interval graphs and their complements. This leads to more natural questions about $k$-DOMINATING SET: Are the two problems $k$-INDEPENDENT DOMINATING SET and $k$-DOMINATING CLIQUE still W[1]-hard in multiple-interval graphs and their complements? Also, without veering to either extreme, how about $k$-CONNECTED DOMINATING SET?

We show that our FPT reduction for the W[1]-hardness of $k$-DOMINATING SET in co-3-track interval graphs in Theorem 2 also establishes the following theorem:

**Theorem 3.** $k$-CONNECTED DOMINATING SET *and* $k$-DOMINATING CLIQUE *in co-3-track interval graphs are both W[1]-hard with parameter $k$.*

Similarly, it is not difficult to verify that the FPT reduction for the W[1]-hardness of $k$-DOMINATING SET in unit 2-track interval graphs [19] also establishes the following theorem:

**Theorem 4.** $k$-INDEPENDENT DOMINATING SET *in unit 2-track interval graphs is W[1]-hard with parameter $k$.*

For the two problems $k$-CONNECTED DOMINATING SET and $k$-DOMINATING CLIQUE in multiple-interval graphs, we obtain a weaker result:

**Theorem 5.** $k$-CONNECTED DOMINATING SET *and* $k$-DOMINATING CLIQUE *in unit 3-track interval graphs are both W[1]-hard with parameter $k$.*

Recall that $k$-DOMINATING SET in general graphs is W[2]-complete. Fellows et al. [11] asked whether it remains W[2]-complete in $t$-interval graphs for $t \geq 2$. Our following theorem shows that this is very unlikely:

**Theorem 6.** $k$-DOMINATING SET, $k$-CONNECTED DOMINATING SET, $k$-INDEPENDENT DOMINATING SET, *and* $k$-DOMINATING CLIQUE *in $t$-interval graphs and co-$t$-interval graphs for all constants $t \geq 2$ are in W[1].*

A generalization of $k$-DOMINATING SET is called $d$-DISTANCE $k$-DOMINATING SET, where each vertex is able to dominate all vertices within a threshold distance $d$. Note that $k$-DOMINATING SET is simply $d$-DISTANCE $k$-DOMINATING SET with $d = 1$. On the other hand, $d$-DISTANCE $k$-DOMINATING SET in any graph $G$ is simply $k$-DOMINATING SET in the $d$th power of $G$. In contrast to Theorems 1 and 6, we have the following theorem for $d$-DISTANCE $k$-DOMINATING SET:

**Theorem 7.** *$d$-DISTANCE $k$-DOMINATING SET for any $d \geq 2$ in unit 2-track interval graphs, for $d = 2$ in co-3-interval graphs, and for any $d \geq 3$ in co-4-interval graphs is W[2]-hard with parameter $k$.*

The last variant of $k$-DOMINATING SET that we study in this paper is called $k$-PERFECT CODE. For a graph $G = (V, E)$ and a vertex $u \in V$, we define the *open neighborhood of $u$ in $G$* as $N(u) := \{v \mid \{u, v\} \in E\}$, and define the *closed neighborhood of $u$ in $G$* as $N[u] := N(u) \cup \{u\}$. A *perfect code* in a graph $G = (V, E)$, also known as a *perfect dominating set* or an *efficient dominating set*, is a subset of vertices $V' \subseteq V$ that includes exactly one vertex from the closed neighborhood of each vertex $u \in V$. The problem $k$-PERFECT CODE is that of deciding whether a given graph $G$ has a perfect code of size exactly $k$.

The problem $k$-PERFECT CODE is W[1]-complete with parameter $k$ in general graphs [8, 4]. It is also known to be NP-complete in $r$-regular graphs for any $r \geq 3$ [23] and in planar graphs of maximum degree 3 [12]. Since every graph of maximum degree 3 is the intersection graph of a family of unit 2-track intervals [20, Theorem 4], it follows that $k$-PERFECT CODE is NP-complete in unit 2-track interval graphs. In the following theorem, we show that $k$-PERFECT CODE is indeed W[1]-hard in unit 2-track interval graphs:

**Theorem 8.** *$k$-PERFECT CODE in unit 2-track interval graphs is W[1]-hard with parameter $k$.*

The distance variant of $k$-PERFECT CODE, denoted as $d$-DISTANCE $k$-PERFECT CODE, is also studied in the literature [23]. Recall that $d$-DISTANCE $k$-DOMINATING SET in any graph $G$ is simply $k$-DOMINATING SET in the $d$th power of $G$. Similarly, $d$-DISTANCE $k$-PERFECT CODE in any graph $G$ is simply $k$-PERFECT CODE in the $d$th power of $G$. Since $k$-PERFECT CODE in general graphs is in W[1] [4], it follows that $d$-DISTANCE $k$-PERFECT CODE in general graphs is also in W[1]. In the following theorem, we show that $d$-DISTANCE $k$-PERFECT CODE is W[1]-hard even in unit 2-track interval graphs:

**Theorem 9.** *$d$-DISTANCE $k$-PERFECT CODE for any $d \geq 2$ in unit 2-track interval graphs is W[1]-hard with parameter $k$.*

At the end of their paper, Fellows et al. [11] listed four problems that are W[1]-complete in general graphs, and suggested that a possibly prosperous direction for extending their work would be to investigate whether these problems become fixed-parameter tractable in multiple-interval graphs. The four problems are $k$-VERTEX CLIQUE COVER, $k$-SEPARATING VERTICES, $k$-PERFECT CODE, and $k$-IRREDUNDANT SET.

The problem $k$-VERTEX CLIQUE COVER has a close relative called $k$-EDGE CLIQUE COVER. Given a graph $G = (V, E)$ and an integer $k$, the problem $k$-VERTEX CLIQUE COVER asks whether the vertex set $V$ can be partitioned into $k$ disjoint subsets $V_i$, $1 \leq i \leq k$, such that each subset $V_i$ induces a complete subgraph of $G$, and the problem $k$-EDGE CLIQUE COVER asks whether there are $k$ (not necessarily disjoint) subsets $V_i$ of $V$, $1 \leq i \leq k$, such that each subset $V_i$ induces a complete subgraph of $G$ and, moreover, for each edge $\{u, v\} \in E$, there is some $V_i$ that contains both $u$ and $v$. The two problems $k$-VERTEX CLIQUE COVER and $k$-EDGE CLIQUE COVER are also known in the literature as $k$-CLIQUE PARTITION and $k$-CLIQUE COVER, respectively, and are both NP-complete [13, GT15 and GT17]. To avoid possible ambiguity, we will henceforth use the term $k$-VERTEX CLIQUE PARTITION instead of $k$-VERTEX CLIQUE COVER or $k$-CLIQUE PARTITION.

Although the two problems $k$-VERTEX CLIQUE PARTITION and $k$-EDGE CLIQUE COVER are both NP-complete, they have very different parameterized complexities. The problem $k$-EDGE CLIQUE COVER is fixed-parameter tractable in general graphs [16]; hence it is also fixed-parameter tractable in multiple-interval graphs and their complements. On the other hand, the problem $k$-VERTEX CLIQUE PARTITION in any graph $G$ is the same as the problem $k$-VERTEX COLORING in the complement graph $\overline{G}$. It is known that 3-VERTEX COLORING of planar graphs of maximum degree 4 is NP-hard [15]. It is also known that $k$-VERTEX COLORING in circular-arc graphs is NP-hard if $k$ is part of the input [14]. Since graphs of maximum degree 4 are unit 3-track interval graphs [20, Theorem 4], and since circular-arc graphs are obviously 2-track interval graphs (by a simple cutting argument), we immediately have the following easy theorem on the complexity of $k$-VERTEX CLIQUE PARTITION in the complements of multiple-interval graphs:

**Theorem 10.** 3-VERTEX CLIQUE PARTITION *in co-unit 3-track interval graphs is NP-hard; thus, unless $NP = P$, $k$-VERTEX CLIQUE PARTITION in co-unit 3-track interval graphs does not admit any FPT algorithms with parameter $k$. Also, $k$-VERTEX CLIQUE PARTITION in co-2-track interval graphs is NP-hard if $k$ is part of the input.*

For the complexity of $k$-VERTEX CLIQUE PARTITION in multiple-interval graphs, we obtain the following theorem:

**Theorem 11.** $k$-VERTEX CLIQUE PARTITION *in unit 2-interval graphs is W[1]-hard with parameter $k$.*

Given a graph $G = (V, E)$ and two integers $k$ and $l$, the problem $k$-SEPARATING VERTICES is that of deciding whether there is a partition $V = X \cup S \cup Y$ of the vertices such that $|X| = l$, $|S| \le k$, and there is no edge between $X$ and $Y$? In other words, is it possible to cut $l$ vertices off the graph by deleting $k$ vertices?

The problem $k$-SEPARATING VERTICES is one of several closely related graph separation problems considered by Marx [24] in terms of parameterized complexity. Marx showed that $k$-SEPARATING VERTICES is W[1]-hard in general graphs with two parameters $k$ and $l$, but is fixed-parameterized tractable with three parameters $k$, $l$, and the maximum degree $d$ of the graph. In the following two theorems, we show that with two parameters $k$ and $l$, $k$-SEPARATING VERTICES remains W[1]-hard in multiple-interval graphs and their complements:

**Theorem 12.** $k$-SEPARATING VERTICES *in balanced 2-track interval graphs is W[1]-hard with parameters $k$ and $l$.*

**Theorem 13.** $k$-SEPARATING VERTICES *in co-balanced 3-track interval graphs is W[1]-hard with parameters $k$ and $l$.*

The problem $k$-SEPARATING VERTICES was studied under the name CUTTING $l$ VERTICES by Marx [24], who also studied two closely related variants called CUTTING $l$ CONNECTED VERTICES and CUTTING INTO $l$ COMPONENTS. In CUTTING $l$ CONNECTED VERTICES, the $l$ vertices that are separated from the rest of $G$ must induce a connected subgraph of $G$. In CUTTING INTO $l$ COMPONENTS, the objective is to delete at most $k$ vertices such that the remaining graph is broken into at least $l$ connected components. Marx showed that CUTTING $l$ CONNECTED VERTICES is W[1]-hard when parameterized by either $k$ or $l$, and is fixed-parameter tractable when parameterized by both $k$ and $l$. We observe that his W[1]-hardness proof with parameter $l$ involves only line graphs, which are obviously a subclass of unit 2-interval graphs. Marx also showed that CUTTING INTO $l$ COMPONENTS is W[1]-hard when parameterized by both $k$ and $l$. In the following two theorems, we extend these W[1]-hardness results to multiple-interval graphs and their complements:

**Theorem 14.** CUTTING $l$ CONNECTED VERTICES *in balanced 2-track interval graphs and co-balanced 3-track interval graphs is W[1]-hard with parameter $k$.*

**Theorem 15.** CUTTING INTO $l$ COMPONENTS *in balanced 2-track interval graphs and co-balanced 3-track interval graphs is W[1]-hard with parameters $k$ and $l$.*

The problem $k$-PERFECT CODE has been covered in Theorems 8 and 9. We now move on to the last problem, $k$-IRREDUNDANT SET. Recall that for a graph $G = (V, E)$, the *open neighborhood* of $u$ is $N(u) = \{v \mid \{u, v\} \in E\}$, and that the *closed neighborhood* of $u$ is $N[u] = N(u) \cup \{u\}$. For a subset $V' \subseteq V$ of vertices, we define the *open neighborhood of $V'$ in $G$* as $N(V') := \cup_{u \in V'} N(u)$ and define the *closed neighborhood of $V'$ in $G$* as $N[V'] := \cup_{u \in V'} N[u]$. An *irredundant set* in a graph $G = (V, E)$ is a subset $V' \subseteq V$ such that each vertex $u \in V'$ is *irredundant*, i.e., $N[V' - \{u\}]$ is a proper subset of $N[V']$. Equivalently, an *irredundant set* in a graph $G = (V, E)$ is a subset $V' \subseteq V$ such that each vertex $u \in V'$ has a *private neighbor* $\pi(u) \in V$ satisfying one of the two following conditions:

1. $\pi(u)$ is adjacent to $u$ but not to any other vertex $v \in V'$.

2. $\pi(u)$ is $u$ itself, and $u$ is not adjacent to any other vertex $v \in V'$. In this case, we say that $u$ is *self-private*.

Note that an independent set is an irredundant set in which every vertex is self-private.

Both $k$-PERFECT CODE and $k$-IRREDUNDANT SET are very important problems in the development of parameterized complexity theory. The problem $k$-PERFECT CODE was shown to be W[1]-hard as early as 1995 [8], but its membership in W[1] was proved much later in 2002 [4]. Indeed this problem was once conjectured by Downey and Fellows [9, p. 487] either to represent an intermediate between W[1] and W[2], or to be complete for W[2]. Similarly, the problem $k$-IRREDUNDANT SET was shown to be in W[1] in 1992 [7], and was once conjectured as an intermediate between FPT and W[1] before it was finally proved to be W[1]-hard in 2000 [10]:

**Theorem 16** (Downey, Fellows, and Raman [10])**.** $k$-IRREDUNDANT SET *in general graphs is W[1]-hard with parameter $k$.*

The celebrated proof of Downey et al. [10] was a major breakthrough in parameterized complexity theory, but it is rather complicated, spanning seven pages. In this paper, we give a very simple alternative proof (less than two pages) of Theorem 16. Our proof is based on an FPT reduction from the W[1]-complete problem $k$-MULTICOLORED CLIQUE [11]: Given a graph $G$ of $n$ vertices and $m$ edges, and a vertex-coloring $\kappa : V(G) \to \{1, 2, \ldots, k\}$, decide whether $G$ has a clique of $k$ vertices containing exactly one vertex of each color (without loss of generality, we assume that no edge in $G$ connects two vertices of the same color). Indeed all proofs of W[1]-hardness in this paper are based on FPT reductions from this problem. After its invention, this technique quickly became a standard tool for parameterized reductions. It was used by researchers to prove new W[1]-hardness results as well as to simplify existing W[1]-hardness proofs in many different settings.

The problem of recognizing multiple-interval graphs is NP-hard in general [20]. This aspect of computational complexity involving the recognition of a class of graphs is quite different from the computational complexities of various optimization problems in such graphs. To avoid confusion, for all optimization problems in multiple-interval graphs and their complements that are studied in this paper, we assume that the multiple-interval representation of the graph is given as part of the input.

## 2 Dominating Set

In this section we prove Theorem 2. We show that $k$-DOMINATING SET in co-3-track interval graphs is W[1]-hard by an FPT reduction from the W[1]-complete problem $k$-MULTICOLORED CLIQUE [11].

Let $(G, \kappa)$ be an instance of $k$-MULTICOLORED CLIQUE. We will construct a family $\mathcal{F}$ of 3-track intervals such that $G$ has a clique of $k$ vertices containing exactly one vertex of each color if and only if the complement of the intersection graph $G_\mathcal{F}$ of $\mathcal{F}$ has a dominating set of $k'$ vertices, where $k' = k + \binom{k}{2}$.

*Vertex selection*: Let $v_1, \ldots, v_n$ be the set of vertices in $G$, sorted by color such that the indices of all vertices of each color are contiguous. For each color $i$, $1 \le i \le k$, let $V_i = \{v_p \mid s_i \le p \le t_i\}$ be the set of vertices $v_p$ of color $i$. For each vertex $v_p$, $1 \le p \le n$, let $\langle v_p \rangle$ be a *vertex 3-track interval* consisting of the following three intervals on the three tracks:

$$\langle v_p \rangle = \begin{cases} \text{track 1}: & (p-1, p) \\ \text{track 2}: & (p-1+m+1, p+m+1) \\ \text{track 3}: & (p-1+m+1, p+m+1). \end{cases}$$

For each color $i$, $1 \le i \le k$, let $\langle V_i \rangle$ be the following 3-track interval:

$$\langle V_i \rangle = \begin{cases} \text{track 1}: & (t_i, m+n+1) \\ \text{track 2}: & (0, s_i - 1 + m + 1) \\ \text{track 3}: & (m, m+1). \end{cases}$$

*Edge selection*: Let $e_1, \ldots, e_m$ be the set of edges in $G$, also sorted by color such that the indices of all edges of each color pair are contiguous. For each pair of distinct colors $i$ and $j$, $1 \le i < j \le k$, let $E_{ij} = \{e_r \mid s_{ij} \le r \le t_{ij}\}$ be the set of edges $v_p v_q$ such that $v_p$ has color $i$ and $v_q$ has color $j$. For each edge $e_r$, $1 \le r \le m$, let $\langle e_r \rangle$ be an *edge 3-track interval* consisting of the following three intervals on the three tracks:

$$\langle e_r \rangle = \begin{cases} \text{track 1}: & (r-1+n+1, r+n+1) \\ \text{track 2}: & (r-1, r) \\ \text{track 3}: & (r-1, r). \end{cases}$$

For each pair of distinct colors $i$ and $j$, $1 \le i < j \le k$, let $\langle E_{ij} \rangle$ be the following 3-track interval:

$$\langle E_{ij} \rangle = \begin{cases} \text{track 1}: & (0, s_{ij} - 1 + n + 1) \\ \text{track 2}: & (t_{ij}, n + m + 1) \\ \text{track 3}: & (m, m+1). \end{cases}$$

*Validation*: For each edge $e_r = v_p v_q$ such that $v_p$ has color $i$ and $v_q$ has color $j$, let $\langle v_p e_r \rangle$ and $\langle v_q e_r \rangle$ be the following 3-track intervals:

$$\langle v_p e_r \rangle = \begin{cases} \text{track 1}: & (p, s_{ij} - 1 + n + 1) \\ \text{track 2}: & (t_{ij}, p - 1 + m + 1) \\ \text{track 3}: & (r - 1, r), \end{cases} \quad \langle v_q e_r \rangle = \begin{cases} \text{track 1}: & (q, s_{ij} - 1 + n + 1) \\ \text{track 2}: & (t_{ij}, q - 1 + m + 1) \\ \text{track 3}: & (r - 1, r). \end{cases}$$

Let $\mathcal{F}$ be the following family of $n + m + k + \binom{k}{2} + 2m$ 3-track intervals:

$$\mathcal{F} = \{\langle v_p \rangle \mid 1 \le p \le n\} \cup \{\langle e_r \rangle \mid 1 \le r \le m\}$$
$$\cup \{\langle V_i \rangle \mid 1 \le i \le k\} \cup \{\langle E_{ij} \rangle \mid 1 \le i < j \le k\}$$
$$\cup \{\langle v_p e_r \rangle, \langle v_q e_r \rangle \mid e_r = v_p v_q \in E_{ij}, 1 \le i < j \le k\}.$$

This completes the construction. We refer to Figure 3 for an example. The following five properties of the construction can be easily verified:

1. For each color $i$, $1 \le i \le k$, all 3-track intervals $\langle v_p \rangle$ for $v_p \in V_i$ are pairwise-disjoint.

2. For each color $i$, $1 \le i \le k$, $\langle V_i \rangle$ intersects all other 3-track intervals except the vertex 3-track intervals $\langle v_p \rangle$ for $v_p \in V_i$.

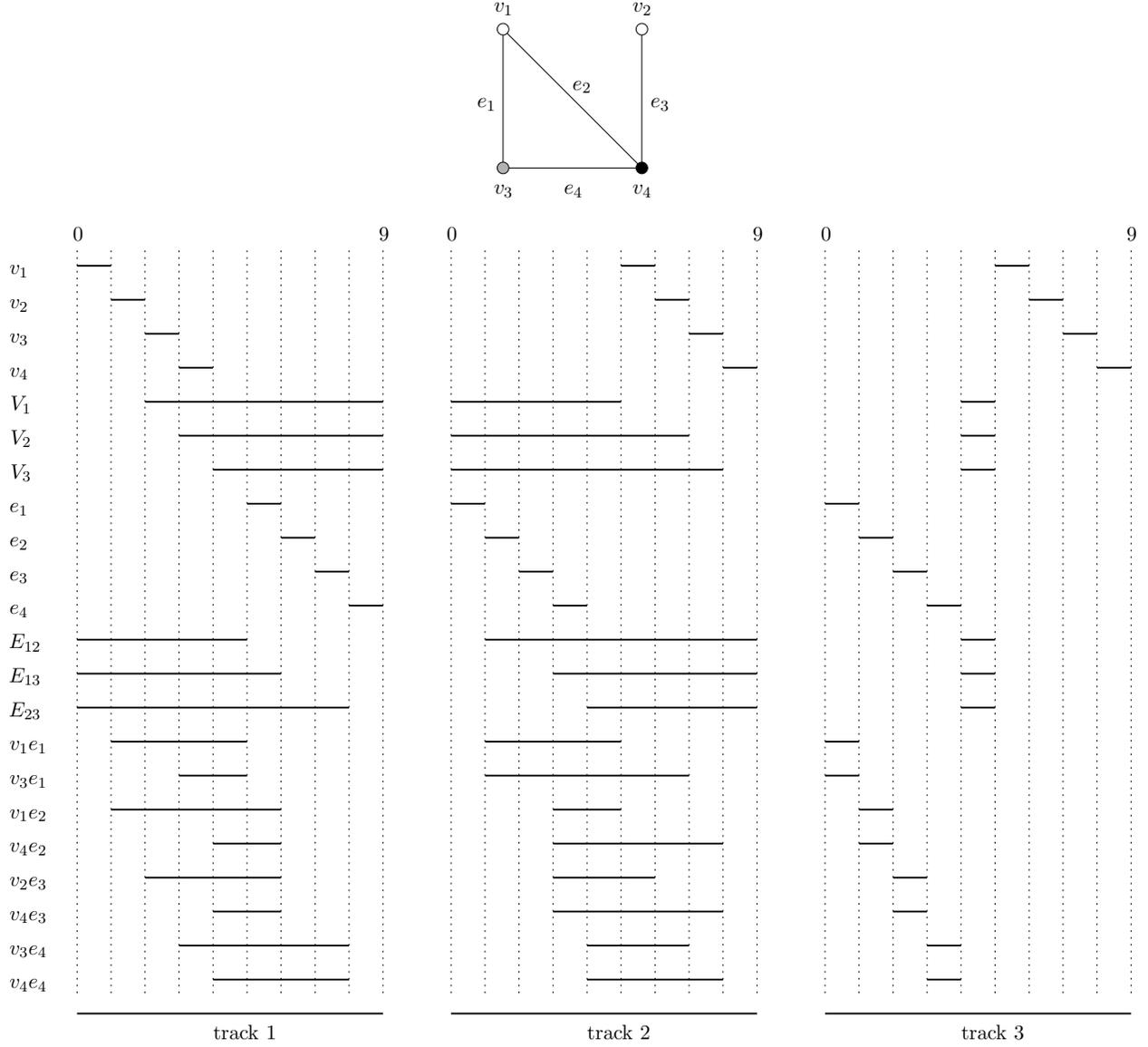

Figure 3: Top: A graph $G$ of $n = 4$ vertices $v_1, v_2, v_3, v_4$ and $m = 4$ edges $e_1 = v_1v_3, e_2 = v_1v_4, e_3 = v_2v_4, e_4 = v_3v_4$, with $k = 3$ colors $\kappa(v_1) = \kappa(v_2) = 1$, $\kappa(v_3) = 2$, and $\kappa(v_4) = 3$. $V_1 = \{v_1, v_2\}, V_2 = \{v_3\}, V_3 = \{v_4\}$; $E_{12} = \{e_1\}, E_{13} = \{e_2, e_3\}, E_{23} = \{e_4\}$. $K = \{v_1, v_3, v_4\}$ is a 3-multicolored clique. Bottom: A family $\mathcal{F}$ of $n+m+k+\binom{k}{2}+2m = 22$ 3-track intervals. $\mathcal{D} = \{\langle v_1 \rangle, \langle v_3 \rangle, \langle v_4 \rangle, \langle e_1 \rangle, \langle e_2 \rangle, \langle e_4 \rangle\}$ is a 6-dominating set in the complement of the intersection graph of $\mathcal{F}$.

3. For each pair of distinct colors $i$ and $j$, $1 \leq i < j \leq k$, all 3-track intervals $\langle e_r \rangle$ for $e_r \in E_{ij}$ are pairwise-disjoint.

4. For each pair of distinct colors $i$ and $j$, $1 \leq i < j \leq k$, $\langle E_{ij} \rangle$ intersects all other 3-track intervals except the edge 3-track intervals $\langle e_r \rangle$ for $e_r \in E_{ij}$.

5. For each pair of distinct colors $i$ and $j$, $1 \leq i < j \leq k$, for each edge $e_r \in E_{ij}$ and each vertex $v_p$ incident to $e_r$, $\langle v_p e_r \rangle$ intersects all other 3-track intervals except the vertex 3-track interval $\langle v_p \rangle$ and the edge 3-track intervals for the edges in $E_{ij}$ other than $\langle e_r \rangle$.

**Lemma 1.** *$G$ has a $k$-multicolored clique if and only if $\overline{G_{\mathcal{F}}}$ has a $k'$-dominating set.*

*Proof.* For the direct implication, if $K \subseteq V(G)$ is a $k$-multicolored clique in $G$, then the following subset $\mathcal{D} \subseteq \mathcal{F}$ of 3-track intervals is a $k'$-dominating set in $\overline{G_{\mathcal{F}}}$:

$$\mathcal{D} = \big\{ \langle v_p \rangle \mid v_p \in K \big\} \cup \big\{ \langle e_r \rangle \mid v_p, v_q \in K,\ e_r = v_p v_q \big\}.$$

To verify this, check that each $\langle v_p \rangle \notin \mathcal{D}$ is dominated by $\langle v_{p'} \rangle \in \mathcal{D}$ for some vertex $v_{p'}$ of the same color as $v_p$ (Property 1), each $\langle e_r \rangle \notin \mathcal{D}$ is dominated by $\langle e_{r'} \rangle \in \mathcal{D}$ for some edge $e_{r'}$ of the same color pair as $e_r$ (Property 3), each $\langle V_i \rangle$ is dominated by $\langle v_p \rangle \in \mathcal{D}$ for some $v_p \in V_i$ (Property 2), each $\langle E_{ij} \rangle$ is dominated by $\langle e_r \rangle \in \mathcal{D}$ for some $e_r \in E_{ij}$ (Property 4), and each $\langle v_p e_r \rangle$ is dominated either by $\langle v_p \rangle \in \mathcal{D}$, when $v_p \in K$, or by $\langle e_{r'} \rangle \in \mathcal{D}$ for some edge $e_{r'}$ of the same color pair as $e_r$, when $v_p \notin K$ (Property 5).

For the reverse implication, suppose that $\mathcal{D} \subseteq \mathcal{F}$ is a $k'$-dominating set in $\overline{G_{\mathcal{F}}}$. We will find a $k$-multicolored clique $K \subseteq V(G)$ in $G$. For each color $i$, $1 \leq i \leq k$, $\mathcal{D}$ must include either $\langle V_i \rangle$ or at least one of its neighbors in $\overline{G_{\mathcal{F}}}$. Thus by Properties 1 and 2, we can assume without loss of generality that $\mathcal{D}$ does not include $\langle V_i \rangle$ but includes at least one vertex 3-track interval $\langle v_p \rangle$ for some $v_p \in V_i$. Similarly, for each pair of distinct colors $i$ and $j$, $1 \leq i < j \leq k$, we can assume by Properties 3 and 4 that $\mathcal{D}$ does not include $\langle E_{ij} \rangle$ but includes at least one edge 3-track interval $\langle e_r \rangle$ for some $e_r \in E_{ij}$. Since $k' = k + \binom{k}{2}$, it follows that $\mathcal{D}$ includes exactly one vertex 3-track interval of each color, and exactly one edge 3-track interval of each color pair. For each pair of distinct colors $i$ and $j$, $1 \leq i < j \leq k$, let $e_r = v_p v_q$ be the edge whose 3-track interval $\langle e_r \rangle$ is included in $\mathcal{D}$. By Property 5 of the construction, the two 3-track intervals $\langle v_p e_r \rangle$ and $\langle v_q e_r \rangle$ cannot be dominated by $\langle e_r \rangle$ and hence must be dominated by $\langle v_p \rangle$ and $\langle v_q \rangle$, respectively. Therefore the vertex selection and the edge selection are consistent, and the set of $k$ vertex 3-track intervals in $\mathcal{D}$ corresponds to a $k$-multicolored clique $K$ in $G$. □

## 3  Connected Dominating Set, Independent Dominating Set, and Dominating Clique

In this section we prove Theorems 3, 4, and 5.

For Theorem 3, to show the W[1]-hardness of $k$-CONNECTED DOMINATING SET and $k$-DOMINATING CLIQUE in co-3-track interval graphs, let us review our FPT reduction for Theorem 2, in particular, the proof of Lemma 1, in the previous section. Observe that for the direct implication of Lemma 1, our proof composes a dominating set $\mathcal{D}$ of pairwise-disjoint 3-track intervals, and that for the reverse implication of Lemma 1, our proof uses only the fact that $\mathcal{D}$ is a dominating set without any assumption about its connectedness. This implies that our FPT reduction for Theorem 2 also establishes Theorem 3. By a similar argument, it is not difficult to verify that the FPT reduction for the W[1]-hardness of $k$-DOMINATING SET in unit 2-track interval graphs [19] also establishes the W[1]-hardness of $k$-INDEPENDENT DOMINATING SET in unit 2-track interval graphs in Theorem 4.

For Theorem 5, to show the W[1]-hardness of $k$-CONNECTED DOMINATING SET and $k$-DOMINATING CLIQUE in unit 3-track interval graphs, we use the same construction as in the previous FPT reduction for the W[1]-hardness of $k$-DOMINATING SET in unit 2-track interval graphs [19] for the first two tracks. Then, on track 3, we use the same (coinciding) unit interval for all multiple-intervals in

$$\mathcal{F}' = \big\{ \widehat{u_i} \mid u \in V_i,\ 1 \leq i \leq k \big\} \cup \big\{ \widehat{u_i v_j}_{\text{left}}, \widehat{u_i v_j}_{\text{right}} \mid uv \in E_{ij},\ 1 \leq i < j \leq k \big\},$$

and use a distinct unit interval disjoint from all other unit intervals for each of the remaining multiple-intervals. Now the dominating set composed in the direct implication of the proof in [19] becomes a clique. Since the reverse implication of the proof in [19] does not depend on the additional intersections between the multiple-intervals in $\mathcal{F}'$, the modified reduction establishes Theorem 5.

# 4 W[1]-membership of Dominating Set and Its Variants

In this section we prove Theorem 6. We show that $k$-DOMINATING SET, $k$-CONNECTED DOMINATING SET, $k$-INDEPENDENT DOMINATING SET, and $k$-DOMINATING CLIQUE in $t$-interval graphs and co-$t$-interval graphs for all constants $t \geq 2$ are in W[1] by FPT reductions to the W[1]-complete problem SHORT TURING MACHINE COMPUTATION [3]. The same problem has been used to prove the W[1]-membership of $k$-PERFECT CODE in general graphs [4] and of $k$-DOMINATING SET in rectangle intersection graphs [25].

We start with two FPT reductions from $k$-DOMINATING SET in $t$-interval graphs and co-$t$-interval graphs, respectively, to SHORT TURING MACHINE COMPUTATION. Let $G_\mathcal{F}$ be the intersection graph of a family $\mathcal{F}$ of $n$ $t$-intervals. Without loss of generality, we assume that the $2nt$ interval endpoints of the $t$-intervals in $\mathcal{F}$ are all distinct. By a standard technique, we can transform any family $\mathcal{I}$ of intervals, in polynomial time, into a family $\mathcal{I}'$ of intervals with distinct endpoints, such that $\mathcal{I}$ and $\mathcal{I}'$ represent the same interval graph.

We first construct a (nondeterministic) Turing machine $M$ that accepts an empty string in $f(k)$ steps for some function $f$ if and only if $G_\mathcal{F}$ has a $k$-dominating set. The crucial observation is the following. Let $\mathcal{D} \subseteq \mathcal{F}$ be a subfamily of $k$ $t$-intervals. Suppose that $\mathcal{D}$ is not a dominating set for $G_\mathcal{F}$. Then there must exist a $t$-interval $I$ in $\mathcal{F} - \mathcal{D}$ that is disjoint from all $t$-intervals in $\mathcal{D}$. Let $P$ be the set of $2kt$ interval endpoints of the $k$ $t$-intervals in $\mathcal{D}$, and let $P' = P \cup \{-\infty, \infty\}$. For the $s$th interval $I_s$ of the $t$-interval $I$, $1 \leq s \leq t$, let $l_s$ be the rightmost point in $P'$ to the left of $I_s$, and let $r_s$ be the leftmost point in $P'$ to the right of $I_s$. Then each pair of points $l_s$ and $r_s$, $1 \leq s \leq t$, specifies a constraint $l_s < I_s < r_s$ on the $t$-interval $I$. The $t$ constraints together form a multiple-interval "range" $I' = (l_1, r_1) \cup \cdots \cup (l_t, r_t)$. Observe that $I \subset I'$ but no $t$-interval $J$ in $\mathcal{D}$ intersects $I'$.

We now describe the reduction. Let $Q$ be the set of $2nt$ interval endpoints of the $n$ $t$-intervals in $\mathcal{F}$, and let $Q' = Q \cup \{-\infty, \infty\}$. Enumerate all combinations $C$ of $t$ constraints based on $Q'$. For each $C$, compute the value of the boolean function $\text{nonempty}(C)$ on whether there exists a $t$-interval $I$ in $\mathcal{F}$ that satisfies $C$. These values will be incorporated directly into the Turing machine as its internal states and transitions. The following is a high-level description of the Turing machine $M$:

1. Guess a subfamily $\mathcal{D} \subseteq \mathcal{F}$ of $k$ $t$-intervals. (This is the only nondeterministic part; the rest of the computation is deterministic.)

2. Let $P$ be the set of $2kt$ interval endpoints of the $k$ $t$-intervals in $\mathcal{D}$, and let $P' = P \cup \{-\infty, \infty\}$. Enumerate all combinations $C$ of $t$ constraints based on $P'$. For each $C$, do the following:

   (a) Check whether there exists a $t$-interval $J$ in $\mathcal{D}$ that intersects the multiple-interval "range" $I'$ formed by $C$.

   (b) If no such $t$-interval $J$ exists, query the precomputed value of the boolean function $\text{nonempty}(C)$. Reject if it is true.

3. Accept.

Recall that $t$ is a constant. With the boolean function $\text{nonempty}(\cdot)$ precomputed and incorporated into the interval states and transitions of the Turing machine $M$, the maximum number of steps of any nondeterministic branch of $M$ is at most $f(k)$ for some function $f$. In particular, it does not depend on $n$ although the size of $M$ itself (i.e., the alphabet size, the number of internal states and transitions, etc.) depends on $n$. Moreover, we can compute $\text{nonempty}(\cdot)$, construct the Turing machine $M$ itself, and compute an upper bound $f(k)$ on the maximum number of steps of $M$, all in time $g(k) \cdot \text{poly}(n)$ for some function $g$. Thus we have an FPT reduction from $k$-DOMINATING SET in $t$-interval graphs to SHORT TURING MACHINE COMPUTATION.

We next construct a (nondeterministic) Turing machine $\overline{M}$ that accepts an empty string in $f(k)$ steps for some function $f$ if and only if $\overline{G_\mathcal{F}}$ has a $k$-dominating set. The crucial observation is the following. Let $\mathcal{D} \subseteq \mathcal{F}$ be a subfamily of $k$ $t$-intervals. Suppose that $\mathcal{D}$ is not a dominating set for $\overline{G_\mathcal{F}}$. Then there must exist a $t$-interval $I$ in $\mathcal{F} - \mathcal{D}$ that intersects all $t$-intervals in $\mathcal{D}$. Let $P$ be the set of $2kt$ interval endpoints of the $k$ $t$-intervals in $\mathcal{D}$, and let $P' = P \cup \{-\infty, \infty\}$. For the $s$th interval $I_s = (p_s, q_s)$ of the $t$-interval $I$, $1 \leq s \leq t$, let $lp_s$ be the rightmost point in $P'$ to the left of $p_s$, let $rp_s$ be the leftmost point in $P'$ to the right of $p_s$, let $lq_s$ be the rightmost point in $P'$ to the left of $q_s$, and let $rq_s$ be the leftmost point in $P'$ to the right of $q_s$. Then each pair of points $lp_s$ and $rp_s$, $1 \leq s \leq t$, specifies a constraint $lp_s < p_s < rp_s$, and each pair of points $lq_s$ and $rq_s$, $1 \leq s \leq t$, specifies a constraint $lq_s < q_s < rq_s$, on the $t$-interval $I$. Let $C$ be this combination of $2t$ constraints. Observe that any $t$-interval $I'$ (not necessarily in $\mathcal{F}$) that satisfies $C$ intersects all $t$-intervals in $\mathcal{D}$.

We now describe the reduction. Let $Q$ be the set of $2nt$ interval endpoints of the $n$ $t$-intervals in $\mathcal{F}$, and let $Q' = Q \cup \{-\infty, \infty\}$. Enumerate all combinations $C$ of $2t$ constraints based on $Q'$. For each $C$, compute the value of the boolean function $\mathrm{nonempty}(C)$ on whether there exists a $t$-interval $I$ in $\mathcal{F}$ that satisfies $C$. These values will be incorporated directly into the Turing machine as its internal states and transitions. The following is a high-level description of the Turing machine $\overline{M}$:

1. Guess a subfamily $\mathcal{D} \subseteq \mathcal{F}$ of $k$ $t$-intervals. (This is the only nondeterministic part; the rest of the computation is deterministic.)

2. Let $P$ be the set of $2kt$ interval endpoints of the $k$ $t$-intervals in $\mathcal{D}$, and let $P' = P \cup \{-\infty, \infty\}$. Sort $P'$. Enumerate all combinations $C$ of $2t$ constraints based on $P'$, subject to the additional condition that the two points in each pair (i.e., the two points $lp_s$ and $rp_s$ in the pair $(lp_s, rp_s)$, or the two points $lq_s$ and $rq_s$ in the pair $(lq_s, rq_s)$, $1 \leq s \leq t$) are consecutive in $P'$. (This additional condition is to ensure that no $t$-interval in $\mathcal{D}$ satisfies $C$.) For each $C$, do the following:

    (a) Check whether there exists a $t$-interval $I'$ (not necessarily in $\mathcal{F}$) that satisfies $C$ and intersects all $t$-intervals in $\mathcal{D}$.

    (b) If such a $t$-interval $I'$ exists, query the precomputed value of the boolean function $\mathrm{nonempty}(C)$. Reject if it is true.

3. Accept.

The analysis is the same as before. Thus we have an FPT reduction from $k$-DOMINATING SET in co-$t$-interval graphs to SHORT TURING MACHINE COMPUTATION.

Finally, to adapt the two reductions to work for the other variants, $k$-CONNECTED DOMINATING SET, $k$-INDEPENDENT DOMINATING SET, and $k$-DOMINATING CLIQUE, it suffices to augment the two Turing machines $M$ and $\overline{M}$ with an additional step that checks whether the subgraph induced by the guessed subfamily $\mathcal{D}$ of $k$ $t$-intervals is connected, is an independent set, and is a clique, respectively.

## 5 Distance Dominating Set

In this section we prove Theorem 7. We show that for any $d \geq 2$ $d$-DISTANCE $k$-DOMINATING SET in multiple-interval graphs and their complements is W[2]-hard by FPT reductions from the W[2]-hard problem $k$-COLORFUL RED-BLUE DOMINATING SET [6]: Given a bipartite graph $G = (R \cup B, E)$ and a vertex-coloring $\kappa : R \to \{1, 2, \ldots, k\}$, decide whether $G$ has a set of $k$ distinctly colored vertices $D \subseteq R$ such that each vertex in $B$ is adjacent to at least one vertex in $D$. We call such a set $D$ a *colorful red-blue dominating set* of $G$.

**Distance Dominating Set in Multiple-Interval Graphs.** First we consider the case $d = 2$. Let $(G, \kappa)$ be an instance of $k$-COLORFUL RED-BLUE DOMINATING SET. We will construct a family $\mathcal{F}$ of 2-track intervals as illustrated in Figure 4.

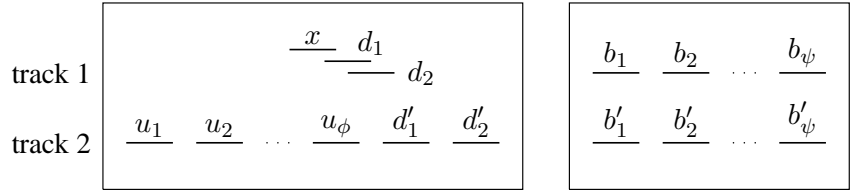

Figure 4: An illustration of the gadgets constructed in the proof of Theorem 7: the gadget for $V_i$ (left) and the gadget for $B$ (right).

For each color $i$, $1 \leq i \leq k$, let $V_i \subseteq R$ be the set of vertices of color $i$. Write $|V_i| = \phi$. We construct $k$ gadgets, one for each $V_i$, $1 \leq i \leq k$. There are three intervals on track 1 labeled with $x, d_1, d_2$. $x$ intersects with $d_1$ and $d_1$ intersects with $d_2$. On track 2, there are $\phi+2$ disjoint intervals labeled with $u_1, \ldots, u_\phi, d'_1, d'_2$. For each vertex $u = u_s \in V_i$, we add a 2-track interval $\langle u \rangle = (x, u_s)$ to $\mathcal{F}$. For each gadget for $V_i$, we also add two dummy 2-track intervals $(d_1, d'_1)$ and $(d_2, d'_2)$ to $\mathcal{F}$.

We then construct one gadget for $B$. Write $|B| = \psi$. Let $b_1, \ldots, b_\psi$ be vertices in $B$. On track 1, there are $\psi$ pairwise disjoint intervals labeled with $b_1 \ldots, b_\psi$. Similarly, on track 2, there are $\psi$ pairwise disjoint intervals labeled with $b'_1, \ldots, b'_\psi$. For each vertex $b = b_t \in B$, add a 2-track interval $\langle b \rangle = (b_t, b'_t)$ to $\mathcal{F}$. Finally, for each edge $e = (u_s, b_t) \in E$ with $u_s \in V_i$ for some $i$ and $b_t \in B$, add a 2-track interval $\langle e \rangle = (b_t, u_s)$ to $\mathcal{F}$. This completes the construction.

In summary, the construction gives us the following family $\mathcal{F}$ of 2-track intervals:

$$\mathcal{F} = \big\{\langle u \rangle \mid u \in V_i,\, 1 \leq i \leq k \big\} \cup \big\{\langle b \rangle \mid b \in B \big\} \cup \big\{\langle e \rangle \mid e \in E \big\} \cup \text{DUMMIES},$$

where DUMMIES is the set of $2k$ dummy 2-track intervals.

**Lemma 2.** *$G$ has a $k$-colorful red-blue dominating set if and only if the intersection graph $G_\mathcal{F}$ of $\mathcal{F}$ has a 2-distance $k$-dominating set.*

*Proof.* We first prove the direct implication. Suppose $G$ has a $k$-colorful red-blue dominating set $K \subseteq R$, then it is easy to verify the family $\mathcal{D} = \big\{\langle u \rangle \mid u \in K \big\}$ of 2-track intervals is a 2-distance $k$-dominating set in $G_\mathcal{F}$.

We next prove the reverse implication. Suppose that $\mathcal{D}$ is a 2-distance $k$-dominating set in $G_\mathcal{F}$. To dominate the two dummy 2-track intervals $(d_1, d'_1)$ and $(d_2, d'_2)$ in the gadget for $V_i$, we can assume without loss of generality that $\mathcal{D}$ includes at least one $\langle u \rangle$ from each gadget for $V_i$. Since $\mathcal{D}$ has size $k$, we must have exactly one $\langle u \rangle$ from each gadget for $V_i$. For any $b \in B$, $\langle b \rangle$ must be dominated by some $\langle u \rangle \in \mathcal{D}$. By the construction, this implies that $(u, b) \in E$. Therefore, the set $\{u \mid \langle u \rangle \in \mathcal{D}\}$ is a $k$-colorful red-blue dominating set for $G$. □

To generalize the above construction to handle the case $d > 2$, it suffices to make only two changes to $G_\mathcal{F}$:

1. For each color $i$, $1 \leq i \leq k$, replace the two dummy vertices by a "path" of $d$ dummy vertices with one end free and one end connected to all vertices in $V_i$.

2. For each vertex $b \in B$, add a "path" of $d-2$ dummy vertices with one end free and one end connected to $b$.

Clearly each dummy vertex can be represented by a unit 2-track interval as before.

**Distance Dominating Set in Complements of Multiple-Interval Graphs.** To show that $d$-DISTANCE $k$-DOMINATING SET is W[2]-hard for $d = 2$ in co-3-interval graphs, we construct a co-3-interval graph $\overline{G_{\mathcal{F}'}}$ which is very similar to $G_{\mathcal{F}}$. We then use the same arguments as in Lemma 2 to show that $G$ has a $k$-colorful red-blue dominating set if and only if $\overline{G_{\mathcal{F}'}}$ has a 2-distance $k$-dominating set.

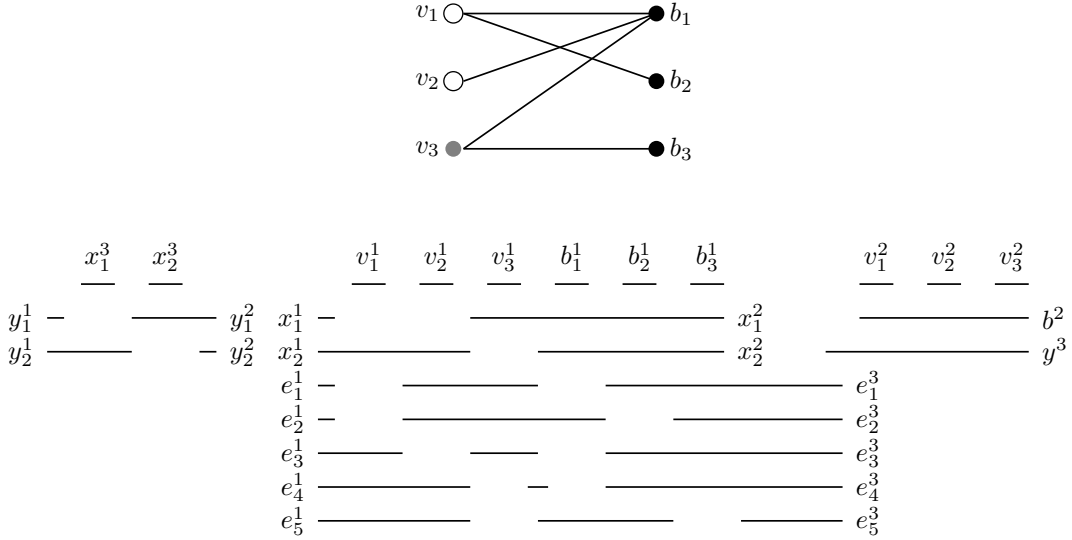

Figure 5: Top: An input graph $G = (R \cup B, E)$ for $k$-COLORFUL RED-BLUE DOMINATING SET, with $R = \{v_1, v_2, v_3\}$, $B = \{b_1, b_2, b_3\}$, and $E = \{e_1 = v_1b_1, e_2 = v_1b_2, e_3 = v_2b_1, e_4 = v_3b_1, e_5 = v_3b_3\}$. There are two color groups $V_1 = \{v_1, v_2\}, V_2 = \{v_3\}$. Bottom: The corresponding construction of $\overline{G_{\mathcal{F}'}}$. Note that the label $e_r^2$ ($1 \leq r \leq 5$), for the interval between $e_r^1$ and $e_r^3$, is omitted.

We briefly describe how $\overline{G_{\mathcal{F}'}}$ is constructed. Refer to Figure 5 for an illustration. For convenience, we specify some 3-intervals in $\mathcal{F}'$ as 2-intervals, and assume an implicit extension of each 2-interval to a 3-interval by adding an extra interval that is disjoint from all other intervals. Given an input graph $G = (R \cup B, E)$ and a vertex-coloring $\kappa : R \to \{1, 2, \ldots, k\}$. Let $v_1, \ldots, v_m$ be an ordering of the vertices in $R$ such that all vertices in any color group $V_i$ are consecutive in the ordering. For each vertex $v_i \in R$, add a 2-interval $(v_i^1, v_i^2)$ to $\mathcal{F}'$. Let $b_1, \ldots, b_n$ be the vertices in $B$. For each vertex $b_j \in B$, add a 2-interval $(b_j^1, b^2)$ to $\mathcal{F}'$. The interval $b^2$ intersects all $v_i^2$. For each edge $e_r = (v_s, b_t) \in E$, add a 3-interval $(e_r^1, e_r^2, e_r^3)$ to $\mathcal{F}'$ such that the three intervals together intersect all $v_i^1$ and $b_j^1$ except $v_s^1$ and $b_t^1$. We then add $k$ dummy 3-intervals $(x_p^1, x_p^2, x_p^3)$, $1 \leq p \leq k$, to $\mathcal{F}'$, such that $x_p^1$ and $x_p^2$ together intersect all $v_i^1$ and $b_j^1$ except those $v_s^1$ for $v_s \in V_p$. The intervals $x_p^3, 1 \leq p \leq k$, are pairwise disjoint. Finally we add $k$ more dummy 3-intervals $(y_q^1, y_q^2, y^3)$, $1 \leq q \leq k$, to $\mathcal{F}'$ such that $y_q^1$ and $y_q^2$ together intersect all $x_p^3$ except $x_q^3$. The interval $y^3$ intersects all $v_i^2$, all $e_r^3$, and $b^2$.

One can check that the intersection graph $\overline{G_{\mathcal{F}'}}$ is almost identical to $G_{\mathcal{F}}$ constructed in Figure 4. The only difference is that in $\overline{G_{\mathcal{F}'}}$ all vertices in $R$ form a big clique whereas in $G_{\mathcal{F}}$ the vertices in each color group $V_i$ form a clique, separately. The arguments in Lemma 2 still apply. Therefore $d$-DISTANCE $k$-DOMINATING SET is W[2]-hard for $d = 2$ in co-3-interval graphs.

Let $G_2 = \overline{G_{\mathcal{F}'}}$ be the co-3-interval graph that we just constructed for $d = 2$. To generalize the above construction to handle the case $d \geq 3$, it suffices to extend the graph $G_2$ to a graph $G_d$ by making the same two changes as before, i.e., adding more dummy vertices. The difficulty now is that for the complements of multiple-interval graphs, three intervals for each vertex are not enough to encode all the edges in the construction. Nevertheless, we show that for $d \geq 3$, four intervals for each vertex are enough. Our proof is by induction. We already have the co-3-interval graph $G_2$ for the base case $d = 2$. Next we consider the inductive step.

For $d = 3$, to obtain $G_3$ from $G_2$, we start with the co-3-interval graph that encodes $G_2$, then extend each dummy path by one more vertex at the free end. Let $R_2$ be the interval region of the real line that contains all 3-intervals in $G_2$. To encode the connection between the new dummy vertices in $G_3$ and the existing vertices in $G_2$, we take an unused interval region $R_3$ of the real line to the right of $R_2$. For each vertex in $G_2$, we place one disjoint interval in $R_3$. For each new dummy vertex in $G_3$, we place two disjoint intervals in $R_3$, to cover all of $R_3$ except the interval for its only neighbor. Thus we have a co-4-interval graph $G_3$ represented by four intervals for each vertex in the subgraph $G_2$ and two intervals for each new dummy vertex in $G_3 - G_2$.

Now, for any $d \geq 4$, to obtain $G_d$ from $G_{d-1}$, we extend the interval region $R_{d-2}$ (to the left when $d$ is even, or the right when $d$ is odd) to a longer interval region $R_d$. To encode the connection between the new dummy vertices in $G_d$ and the existing vertices in $G_{d-1}$, we place one disjoint interval in $R_d - R_{d-2}$ for each dummy vertex in $G_{d-1} - G_{d-2}$, and place two disjoint intervals in $R_d$ for each new dummy vertex in $G_d - G_{d-1}$, to cover all of $R_d$ except the interval in $R_d - R_{d-2}$ for its only neighbor in $G_{d-1} - G_{d-2}$. Thus we have a co-4-interval graph $G_d$ represented by at most four intervals for each vertex of the subgraph $G_{d-1}$ and two intervals for each new dummy vertex in $G_d - G_{d-1}$.

## 6 Perfect Code

In this section we prove Theorem 8. We show that $k$-PERFECT CODE in unit 2-track interval graphs is W[1]-hard by a reduction from $k$-MULTICOLORED CLIQUE.

Let $(G, \kappa)$ be an instance of $k$-MULTICOLORED CLIQUE. We will construct a family $\mathcal{F}$ of unit 2-track intervals such that $G$ has a $k$-multicolored clique if and only if the intersection graph $G_\mathcal{F}$ of $\mathcal{F}$ has a $k'$-perfect code, where $k' = k + 2\binom{k}{2}$.

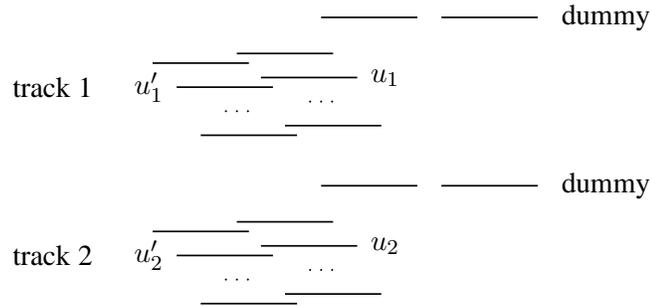

Figure 6: An illustration of a vertex-selection gadget.

*Vertex selection*: For each color $i$, $1 \leq i \leq k$, let $V_i$ be the set of vertices of color $i$. We construct a vertex-selection gadget for $V_i$ as illustrated in Figure 6. Write $|V_i| = \phi$. On each track, we start with $2\phi$ unit intervals arranged in $\phi$ rows and two (slanted) columns. The $\phi$ intervals in each column are pairwise-intersecting. The two intervals in each row slightly overlap such that each interval in the left column intersects with all intervals in the same or higher rows in the right column. For the $r$th vertex $u$ in $V_i$, $1 \leq r \leq \phi$, we add a *vertex 2-track interval* $\langle u \rangle = (u_1, u_2)$ to $\mathcal{F}$, where $u_1$ and $u_2$ are the intervals in the $r$th row and the right column on tracks 1 and 2, respectively. Denote by $u'_1$ and $u'_2$ the intervals in the $r$th row and the left column on tracks 1 and 2, respectively; they will be used for validation. Besides the $\phi$ vertex 2-track intervals $\langle u \rangle$, we also add two dummy 2-track intervals to $\mathcal{F}$. The first (resp. second) dummy 2-interval consists of a unit interval on track 1 (resp. track 2) that intersects all intervals in the right column and no interval in the left column, and a unit interval on track 2 (resp. track 1) that is disjoint from all other intervals.

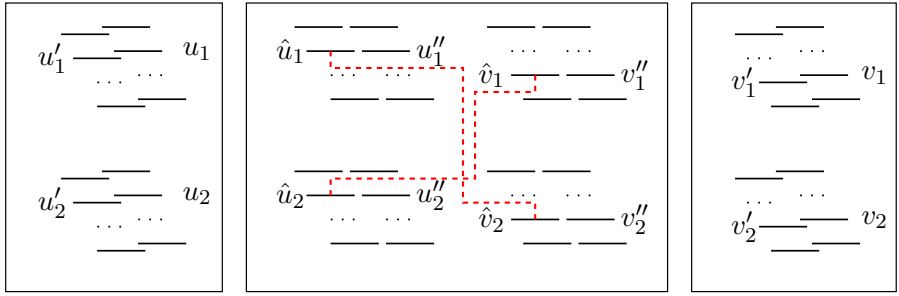

Figure 7: An illustration of an edge-selection gadget (middle) and the corresponding vertex-selection gadgets (left and right). Two edge 2-track intervals $(\hat{u}_1, \hat{v}_2)$ and $(\hat{u}_2, \hat{v}_1)$ are represented by dashed lines. Dummy 2-track intervals are omitted from the figure.

*Edge selection*: For each pair of distinct colors $i$ and $j$, $1 \leq i < j \leq k$, let $E_{ij}$ be the set of edges $uv$ such that $u$ has color $i$ and $v$ has color $j$. We construct an edge selection gadget for $E_{ij}$ as illustrated in Figure 7. We start with four disjoint groups of intervals, two groups on each track, with two columns of intervals in each group. Write $|V_i| = \phi_i$ and $|V_j| = \phi_j$. The two groups on the left correspond to color $i$ and have $\phi_i$ rows; the two groups on the right correspond to color $j$ and have $\phi_j$ rows. Different from the formation in the vertex selection gadgets, here in each group each interval in the left column intersects with all intervals in higher rows in the right column but not the interval in the same row. In the two groups on the left, for the $r$th vertex $u \in V_i$, $1 \leq r \leq \phi_i$, denote by $\hat{u}_1$ and $\hat{u}_2$ the intervals in the $r$th row and the left column on tracks 1 and 2, respectively, and denote by $u_1''$ and $u_2''$ the intervals in the $r$th row and the right column on tracks 1 and 2, respectively. Similarly, for each vertex $v \in V_j$, denote by $\hat{v}_1, \hat{v}_2, v_1', v_2'$ the corresponding intervals in the two groups on the right. For each edge $uv \in E_{ij}$, we add two *edge 2-track intervals* $\langle uv \rangle_1 = (\hat{u}_1, \hat{v}_2)$ and $\langle uv \rangle_2 = (\hat{u}_2, \hat{v}_1)$ to $\mathcal{F}$. Besides these edge 2-track intervals, we also add four dummy 2-track intervals to $\mathcal{F}$, one for each group of intervals. The dummy 2-track interval for each group consists of a unit interval that intersects all intervals in the left column and no interval in the right column in the group, and a unit interval on the other track that is disjoint from all other intervals.

*Validation*: For each pair of distinct colors $i$ and $j$, $1 \leq i < j \leq k$, we add $2|V_i| + 2|V_j|$ *validation 2-track intervals* to $\mathcal{F}$ as illustrated in Figure 7. Specifically, for each vertex $u \in V_i$, we add $\langle u*_{ij} \rangle_1 = (u_1', u_2'')$ and $\langle u*_{ij} \rangle_2 = (u_2', u_1'')$, and for each vertex $v \in V_j$, we add $\langle *v_{ij} \rangle_1 = (v_1', v_2'')$ and $\langle *v_{ij} \rangle_2 = (v_2', v_1'')$.

In summary, the construction gives us the following family $\mathcal{F}$ of unit 2-track intervals:

$$\mathcal{F} = \big\{ \langle u \rangle \mid u \in V_i, 1 \leq i \leq k \big\} \cup \big\{ \langle uv \rangle_1, \langle uv \rangle_2 \mid uv \in E_{ij}, 1 \leq i < j \leq k \big\}$$
$$\cup \big\{ \langle u*_{ij} \rangle_1, \langle u*_{ij} \rangle_2, \langle *v_{ij} \rangle_1, \langle *v_{ij} \rangle_2 \mid u \in V_i, v \in V_j, 1 \leq i < j \leq k \big\} \cup \text{DUMMIES},$$

where DUMMIES is the set of $2k + 4\binom{k}{2}$ dummy 2-track intervals.

**Lemma 3.** *$G$ has a $k$-multicolored clique if and only if $G_\mathcal{F}$ has a $k'$-perfect code.*

*Proof.* We first prove the direct implication. Suppose $G$ has a $k$-multicolored clique $K \subseteq V(G)$, then it is easy to verify that the following subfamily $\mathcal{D}$ of unit 2-track intervals is a $k'$-perfect code in $G_\mathcal{F}$:

$$\mathcal{D} = \big\{ \langle u \rangle \mid u \in K \big\} \cup \big\{ \langle uv \rangle_1, \langle uv \rangle_2 \mid u, v \in K \big\}.$$

We next prove the reverse implication. Suppose $\mathcal{D}$ is a $k'$-perfect code in $G_{\mathcal{F}}$. Observe that the dummy 2-track intervals in our construction are pairwise-disjoint. Moreover, the two dummies in each vertex gadget share the same open neighborhood which is not empty, and the same is true about the two dummies associated with the two groups of intervals, the left group on track 1 and the right group on track 2 (resp. the right group on track 1 and the left group on track 2) of each edge gadget. It follows that these dummies cannot be included in $\mathcal{D}$. In order to perfectly dominate the dummies, $\mathcal{D}$ must include exactly one vertex 2-track interval $\langle u \rangle$ from each vertex selection gadget and two edge 2-track intervals $\langle uv \rangle_1$ and $\langle xy \rangle_2$ from each edge selection gadget. Consider an edge 2-track interval $\langle uv \rangle_1 = (\hat{u}_1, \hat{v}_2)$ from the edge selection gadget for $E_{ij}$, and observe the validation 2-track intervals dominated by $\langle uv \rangle_1$. To perfectly dominate the validation 2-track intervals $\langle w*_{ij} \rangle_2$ for all $w \in V_i$, $\mathcal{D}$ must include $\langle u \rangle$ from the vertex selection gadget for $V_i$. Similarly, to perfectly dominate the validation 2-track intervals $\langle *w_{ij} \rangle_1$ for all $w \in V_j$, $\mathcal{D}$ must include $\langle v \rangle$ from the vertex selection gadget for $V_j$. Then, to perfectly dominate the validation 2-track intervals $\langle w*_{ij} \rangle_1$ for all $w \in V_i$, and $\langle *w_{ij} \rangle_2$ for all $w \in V_j$, the two intervals $\hat{u}_2$ and $\hat{v}_1$ must be used. This implies that the other edge 2-track interval from the same edge selection gadget must be $\langle uv \rangle_2 = (\hat{u}_2, \hat{v}_1)$. Therefore the subset of vertices $K = \{u \in V(G) \mid \langle u \rangle \in \mathcal{D}\}$ is a $k$-multicolored clique in $G$. □

## 7 Distance Perfect Code

In this section we prove Theorem 9. We show that for any $d \geq 2$ $d$-DISTANCE $k$-PERFECT CODE is W[1]-hard in unit 2-interval graphs by FPT reductions from $k$-MULTICOLORED CLIQUE.

We consider the case $d = 2$ first. Let $(G, \kappa)$ be an instance of $k$-MULTICOLORED CLIQUE. We will construct a family $\mathcal{F}$ of unit 2-intervals as illustrated in Figure 8 such that $G$ has a $k$-multicolored clique if and only if the intersection graph $G_{\mathcal{F}}$ of $\mathcal{F}$ has a 2-distance $k'$-perfect code, where $k' = k + \binom{k}{2}$.

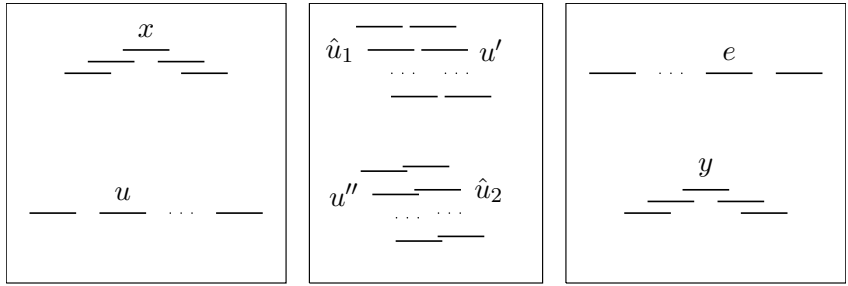

Figure 8: The vertex gadget for $V_i$ (left) is connected to the edge gadget for $E_{ij}$ (right) by a validation gadget (middle).

*Vertex selection*: For each color $i$, $1 \leq i \leq k$, let $V_i$ be the set of vertices of color $i$. We construct a vertex-selection gadget for $V_i$ as illustrated in Figure 8. Write $|V_i| = \phi$. On track 1 there is an interval labeled by $x$. On track 2 there are $\phi$ disjoint intervals, one for each vertex in $V_i$. For the $r$th vertex $u$ in $V_i$, $1 \leq r \leq \phi$, we add a 2-track interval $\langle u \rangle = (x, u)$ to $\mathcal{F}$. We also add four dummy 2-track intervals to $\mathcal{F}$: two dummy 2-track intervals intersect with $x$; the other two dummy 2-track intervals intersect with the first two dummy 2-track intervals, respectively. In figure 8, only one interval (on track 1) of each dummy 2-track intervals is drawn.

*Edge selection*: For each pair of distinct colors $i$ and $j$, $1 \leq i < j \leq k$, let $E_{ij}$ be the set of edges $uv$ such that $u$ has color $i$ and $v$ has color $j$. Write $|E_{ij}| = \psi$. There are $\psi$ disjoint intervals on track 1, one for each edge in $E_{ij}$. There is an interval labeled by $y$ on track 2. For each edge $e \in E_{ij}$, add a 2-track interval $\langle e \rangle = (y, e)$ to $\mathcal{F}$. We also add four dummy 2-track intervals to $\mathcal{F}$ in the similar way as in each vertex selection gadget.

*Validation selection*: For each pair of distinct colors $i$ and $j$, $1 \le i < j \le k$, we construct two validation gadgets that connect the two vertex gadgets for $V_i$ and $V_j$, respectively, to the edge gadget for $E_{ij}$. First we describe the validation gadget between the vertex gadget for $V_i$ and the edge gadget for $E_{ij}$. Write $|V_i| = \phi$ and $|E_{ij}| = \psi$. On track 1, there are $2\phi$ interval arranged in $\phi$ rows and two (slanted) columns. The $\phi$ intervals in each column are pairwise-intersecting. Moreover, each interval in the left column intersects with all intervals in higher rows in the right column but not the interval in the same row. For the $r$th vertex $u \in V_i$, $1 \le r \le \phi$, denote by $\hat{u}_1$ and $u'$ the left and right intervals, respectively, in the $r$th row. On track 2, the arrangement of the $2\phi$ intervals are similar except that each interval in the left column intersects with all intervals in the higher rows *and* the interval in the same row. Denote by $u''$ and $\hat{u}_2$ the left and right intervals, respectively, in the $r$th row. We add $2\phi + \psi$ validation 2-track intervals to $\mathcal{F}$. For each vertex $u \in V_i$, add $\langle u*_{ij}\rangle_1 = (u, u')$ and $\langle u*_{ij}\rangle_2 = (\hat{u}_1, \hat{u}_2)$ to $\mathcal{F}$. For each edge $e = uv \in E_{ij}$, add $\langle u, e\rangle = (e, u'')$ to $\mathcal{F}$.

The validation gadget between the vertex gadget for $V_j$ and the edge gadget for $E_{ij}$ (not shown in Figure 8) is constructed similarly. For each vertex $v \in V_j$, we add $\langle *v_{ij}\rangle_1 = (v, v')$ and $\langle *v_{ij}\rangle_2 = (\hat{v}_1, \hat{v}_2)$ to $\mathcal{F}$. For each edge $e = uv \in E_{ij}$, we add $\langle v, e\rangle = (e, v'')$ to $\mathcal{F}$.

In summary, the construction gives us the following family $\mathcal{F}$ of unit 2-track intervals:

$$\mathcal{F} = \{\langle u\rangle \mid u \in V_i, 1 \le i \le k\} \cup \{\langle e\rangle \mid e \in E_{ij}, 1 \le i < j \le k\}$$
$$\cup \{\langle u*_{ij}\rangle_1, \langle u*_{ij}\rangle_2, \langle *v_{ij}\rangle_1, \langle *v_{ij}\rangle_2 \mid u \in V_i, v \in V_j, 1 \le i < j \le k\}$$
$$\cup \{\langle u, e\rangle, \langle v, e\rangle \mid e = uv \in E_{ij}, 1 \le i < j \le k\} \cup \text{DUMMIES},$$

where DUMMIES is the set of $4k + 4\binom{k}{2}$ dummy 2-track intervals.

**Lemma 4.** *$G$ has a $k$-multicolored clique if and only if $G_\mathcal{F}$ has a 2-distance $k'$-perfect code.*

*Proof.* We first prove the direct implication. Suppose $G$ has a $k$-multicolored clique $K \subseteq V(G)$, then one can verify that the following subfamily $\mathcal{D}$ of 2-track intervals is a 2-distance $k'$-perfect code in $G_\mathcal{F}$:

$$\mathcal{D} = \{\langle u\rangle \mid u \in K\} \cup \{\langle e\rangle \mid e = uv, u, v \in K\}.$$

We next prove the reverse implication. Suppose that $\mathcal{D}$ is a 2-distance $k'$-perfect code in $G_\mathcal{F}$. By a similar argument as in the proof of Lemma 3, the dummies cannot be included in $\mathcal{D}$. In order to perfectly dominate the dummies, $\mathcal{D}$ must include exactly one $\langle u\rangle$ from each vertex gadget and exactly one $\langle e\rangle$ from each edge gadget. For the $r$th vertex $u$ and $t$th vertex $w$ in $V_i$, we write $u \le_i w$ if $r \le t$ and $u >_i w$ if $r > t$. Consider $\langle e\rangle$ from the edge gadget for $E_{ij}$, where $e = uv$. Observe that in the validation gadget between the vertex gadget for $V_i$ and the edge gadget for $E_{ij}$, the 2-track intervals $\{\langle w*_{ij}\rangle_2 \mid w \in V_i, w \le_i u\}$ are within distance 2 from $\langle e\rangle$. Then, to perfectly dominate the 2-track intervals $\{\langle w*_{ij}\rangle_2 \mid w \in V_i, w >_i u\}$, the 2-track interval $\langle u\rangle$ from the vertex gadget for $V_i$ must be included in $\mathcal{D}$. Similarly, to perfectly dominate the 2-track intervals $\langle *w_{ij}\rangle_2$ in the other validation gadget, the 2-track interval $\langle v\rangle$ from the vertex gadget for $V_j$ must also be included in $\mathcal{D}$. Therefore the subset of vertices $K = \{u \in V(G) \mid \langle u\rangle \in \mathcal{D}\}$ is a $k$-multicolored clique in $G$. □

The above construction can be generalized to handle the case $d > 2$. The generalizations for even and odd $d$ are slightly different. We first describe the generalization for even $d$. Extend each vertex gadget to include $d$ pairs of dummy 2-track intervals instead of two pairs, and to include $d-1$ disjoint intervals for each vertex $u$, labeled by $u_s$, $1 \le s \le d-1$, where $u_s$ is on track 2 for odd $s$ and on track 1 for even $s$. Instead of two 2-track intervals $(x, u)$ and $(u, u')$, $d$ 2-track intervals $(x, u_1), (u_1, u_2), \ldots, (u_{d-2}, u_{d-1}), (u_{d-1}, u')$ are added to $\mathcal{F}$. Extend each edge gadget in a similar way to include $d$ pairs of dummy 2-track intervals, and to include $d-1$ disjoint intervals for each edge $e$, labeled by $e_s$, $1 \le s \le d-1$, where $e_s$ is on track 1 for odd $s$ and on track 2 for even $s$. Instead of $(y, e)$ and $(e, u'')$, we have $(y, e_1), (e_1, e_2), \ldots, (e_{d-2}, e_{d-1})$, $(e_{d-1}, u'')$. The generalization for odd $d$ is the same as the generalization for even $d$ except that for each validation gadget we need to swap the intervals on the two tracks, to ensure that $(u_{d-1}, u')$, $(v_{d-1}, v')$, $(e_{d-1}, u'')$, and $(e_{d-1}, v'')$ are indeed 2-track intervals.

## 8 Vertex Clique Partition

In this section we prove Theorem 11. We show that $k$-VERTEX CLIQUE PARTITION in unit 2-interval graphs is W[1]-hard by an FPT reduction from the W[1]-complete problem $k$-MULTICOLORED CLIQUE [11].

Let $(G, \kappa)$ be an instance of $k$-MULTICOLORED CLIQUE. We will construct a family $\mathcal{F}$ of unit 2-intervals such that $G$ has a clique of $k$ vertices containing exactly one vertex of each color if and only if the vertices of the intersection graph $G_\mathcal{F}$ of $\mathcal{F}$ can be partitioned into $k'$ cliques, where $k' = 3k + 2\binom{k}{2}$.

Denote by $C_n$ the cycle graph of $n$ vertices $c_1, \ldots, c_n$ and $n$ edges $c_i c_{i+1}$, $1 \le i \le n-1$, and $c_n c_1$. We first prove the following technical lemma:

**Lemma 5.** *For each integer $n \ge 1$, the cycle graph $C_{4n+1}$ satisfies the following properties:*

1. *The chromatic number of $C_{4n+1}$ is 3.*

2. *The chromatic number of the graph obtained from $C_{4n+1}$ by deleting at least 1 and at most $2n$ vertices, is 2.*

3. *In any partition of the vertices of $C_{4n+1}$ into 3 independent sets, at most one independent set can have size one.*

4. *The complement graph $\overline{C_{4n+1}}$ is a unit 2-interval graph. Moreover, there exists a 2-partition $A_n \cup B_{3n+1}$ of the vertices such that the graph can be represented by one unit interval for each vertex $a_i \in A_n$, $1 \le i \le n$, and two unit intervals for each vertex $b_j \in B_{3n+1}$, $1 \le j \le 3n+1$.*

*Proof.* We prove the four properties one by one:

1. $C_{4n+1}$ is an odd cycle; hence it is not bipartite and has chromatic number at least 3. To achieve the chromatic number 3, we can assign each vertex $c_i$ the color 1 if $i$ is odd but not equal to $4n+1$, the color 2 if $i$ is even, and the color 3 if $i$ is equal to $4n+1$.

2. With any vertex deleted from $C_{4n+1}$, the resulting graph does not have any cycles and hence is bipartite, with chromatic number at most 2. Note that the number of edges in $C_{4n+1}$ is $4n+1$, and that each vertex is incident to 2 edges. With at most $2n$ vertices deleted from $C_{4n+1}$, the resulting graph has at least one edge remaining, and hence has chromatic number at least 2.

3. Let $I_1 \cup I_2 \cup I_3$ be any partition of the vertices of $C_{4n+1}$ into 3 independent sets. Again note that the number of edges in $C_{4n+1}$ is $4n+1 \ge 5$, and that each vertex is incident to 2 edges. If both $I_1$ and $I_2$ have size one, then the 2 vertices in $I_1 \cup I_2$ are together incident to at most 4 edges, and there must be at least one edge remaining between two vertices in $I_3$, which contradicts our assumption that it is an independent set.

4. Consider $4n+1$ vertices spread evenly on a circle of unit perimeter. Connect each vertex to the two farthest vertices by two edges. Then we obtain the cycle graph $C_{4n+1}$. The complement graph $\overline{C_{4n+1}}$ is clearly a circular-arc graph, i.e., the intersection graph of a set of circular-arcs, where each vertex is represented by an open circular arc of length $\frac{2n}{4n+1}$. Let $A_n$ be any $n$ consecutive vertices along the circle and let $B_{3n+1}$ be the remaining $3n+1$ vertices. Then the circular-arc representation of $\overline{C_{4n+1}}$ can be easily "cut" and "stretched" into a 2-interval representation, with one unit interval for each vertex $a_i \in A_n$, $1 \le i \le n$, and with two unit intervals for each vertex $b_j \in B_{3n+1}$, $1 \le j \le 3n+1$. We refer to Figure 9 for an example with $n = 3$.

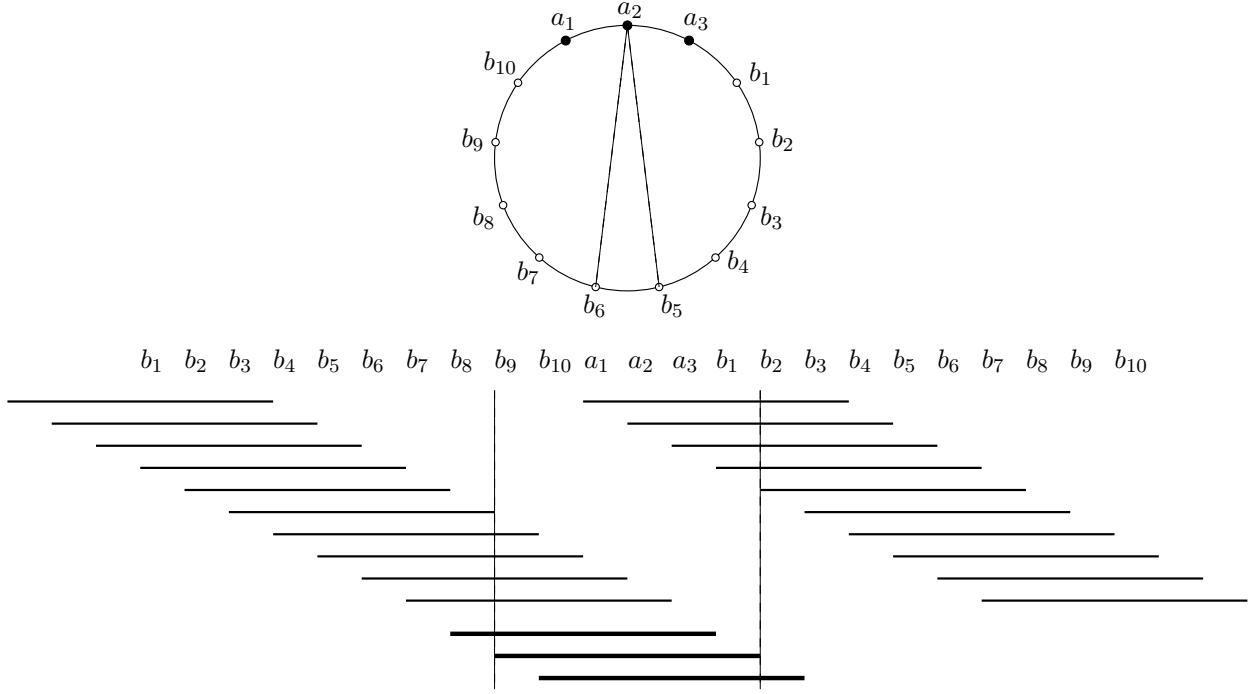

Figure 9: Let $a_1 \ldots a_n b_1 \ldots b_{3n+1}$ be the $4n+1$ vertices along the circle. Then $\overline{C_{4n+1}}$ can be represented by one unit interval for each $a_i$ and two unit intervals for each $b_j$ in the order $b_1 \ldots b_{3n+1} a_1 \ldots a_n b_1 \ldots b_{3n+1}$.

$\square$

*Vertex selection*: Refer to Figure 10(a). For each color $i$, $1 \le i \le k$, let $V_i$ be the set of vertices of color $i$. Let $n_i = |V_i|$. Construct a graph $\overline{C_{4n_i+1}}$ on the $n_i$ vertices in $V_i$ and $3n_i + 1$ additional dummy vertices, represented (using Property 4) by one unit interval for each vertex in $V_i$, and two unit intervals for each dummy vertex. This leaves one free interval for each vertex in $V_i$. Put these $n_i$ free intervals aside, pairwise-disjoint. Thus we have $n_i$ unit 2-intervals including one unit 2-interval $\langle u \rangle$ for each vertex $u \in V_i$, and $3n_i + 1$ additional dummy unit 2-intervals.

*Edge selection*: Refer to Figure 10(b). For each pair of distinct colors $i$ and $j$, $1 \le i < j \le k$, let $E_{ij}$ be the set of edges $uv$ such that $u$ has color $i$ and $v$ has color $j$. Let $m_{ij} = |E_{ij}|$. Construct a graph $\overline{C_{4m_{ij}+1}}$ on $m_{ij}$ vertices (one for each edge in $E_{ij}$) and $3m_{ij} + 1$ additional dummy vertices, represented (using Property 4) by one unit interval for each vertex that corresponds to an edge in $E_{ij}$, and two unit intervals for each dummy vertex. For each edge $uv = e \in E_{ij}$, we construct two unit 2-intervals $\langle ue \rangle$ and $\langle ve \rangle$. Let $\langle e \rangle$ be the unit interval in the representation of $\overline{C_{4m_{ij}+1}}$ that corresponds to the edge $e$. The two unit 2-intervals $\langle ue \rangle$ and $\langle ve \rangle$ share $\langle e \rangle$ as one unit interval, and each of them has one more free interval. Thus we have $2m_{ij}$ unit 2-intervals including two unit 2-intervals $\langle ue \rangle$ and $\langle ve \rangle$ for each edge $uv = e \in E_{ij}$, and $3m_{ij} + 1$ additional dummy unit 2-intervals.

*Validation*: Refer to Figure 10(c). For each edge $uv = e \in E_{ij}$, place the free interval of $\langle ue \rangle$ to coincide with the free interval of $\langle u \rangle$, and place the free interval of $\langle ve \rangle$ to coincide with the free interval of $\langle v \rangle$.

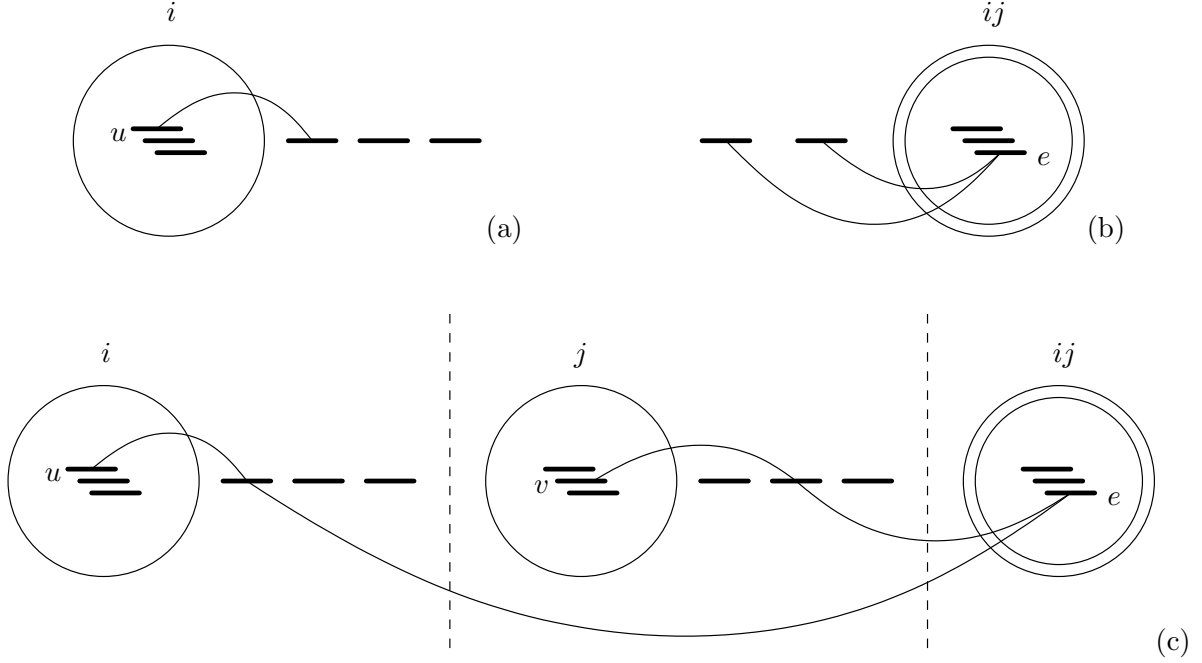

Figure 10: An illustration of the construction for $k$-VERTEX CLIQUE PARTITION. (a) Vertex selection. (b) Edge selection. (c) Validation.

Let $\mathcal{F}$ be the following family of $n + 2m + (3n + 3m + k + \binom{k}{2})$ unit 2-intervals:

$$\mathcal{F} = \big\{\langle u \rangle \mid u \in V_i,\, 1 \le i \le k\big\} \cup \big\{\langle ue \rangle, \langle ve \rangle \mid uv = e \in E_{ij},\, 1 \le i < j \le k\big\} \cup \text{DUMMIES},$$

where DUMMIES is the set of $\sum_i (3n_i + 1) + \sum_{ij}(3m_{ij} + 1) = 3n + 3m + k + \binom{k}{2}$ dummy unit 2-intervals. This completes the construction.

**Lemma 6.** *$G$ has a $k$-multicolored clique if and only if $G_\mathcal{F}$ has a $k'$-vertex clique partition.*

*Proof.* We first prove the direct implication. Suppose that $G$ has a $k$-multicolored clique $K$. We partition $G_\mathcal{F}$ into $k' = 3k + 2\binom{k}{2}$ cliques as follows:

- For each color $i$, $1 \le i \le k$, let $S_i$ be the subgraph of $G_\mathcal{F}$ represented by the $4n_i + 1$ 2-intervals for the $n_i$ vertices in $V_i$ and the $3n_i + 1$ additional dummy vertices. Let $u_i$ be the vertex of color $i$ in $K$. Put the 2-interval $\langle u_i \rangle$, together with the 2-intervals $\langle u_i e \rangle$ for all edges $e$ incident to $u_i$, into one clique. Since $S_i$ is isomorphic to $\overline{C_{4n_i+1}}$, it follows by Property 2 that the remaining $4n_i$ 2-intervals in $S_i$ can be partitioned into two cliques. Thus we have three cliques for each color.

- For each pair of distinct colors $i$ and $j$, $1 \le i < j \le k$, let $S_{ij}$ be the subgraph of $G_\mathcal{F}$ represented by the $5m_{ij} + 1$ 2-intervals including the two 2-intervals $\langle ue \rangle$ and $\langle ve \rangle$ for each edge $uv = e \in E_{ij}$ and the $3m_{ij} + 1$ additional dummy vertices. Let $S'_{ij}$ be the graph obtained from $S_{ij}$ by contracting each pair of vertices represented by $\langle ue \rangle$ and $\langle ve \rangle$ for some edge $e$ (they have the same open neighborhood in $S_{ij}$) into a single vertex represented by $\langle e \rangle$. Then $S'_{ij}$ is isomorphic to $\overline{C_{4m_{ij}+1}}$. Let $u_i v_j = e_{ij}$ be the edge in $K$ such that $u_i$ has color $i$ and $v_j$ has color $j$. The two 2-intervals $\langle u_i e_{ij} \rangle$ and $\langle v_i e_{ij} \rangle$ have already been included in the two cliques containing $\langle u_i \rangle$ and $\langle v_i \rangle$, respectively. Excluding $\langle e_{ij} \rangle$, the remaining $4m_{ij}$ 2-intervals in $S'_{ij}$ can be partitioned into two cliques by Property 2. Now expand each contracted vertex back into two vertices. The two cliques in $S'_{ij}$ remain two cliques in $S_{ij}$. Thus we have two cliques for each pair of distinct colors.

We next prove the reverse implication. Suppose that $G_\mathcal{F}$ has a $k'$-vertex clique partition. We will find a $k$-multicolored clique in $G$. Define the subgraphs $S_i$, $1 \le i \le k$, and the subgraphs $S_{ij}$ and $S'_{ij}$, $1 \le i < j \le k$, as before. By Property 1, each subgraph $S_i$ of $G_\mathcal{F}$ can be partitioned into 3 but no less than 3 cliques. Define $S''_{ij}$, $1 \le i < j \le k$, as the subgraph of $S_{ij}$ (and of $S'_{ij}$) induced by the $3m_{ij} + 1$ dummy vertices. Since $S''_{ij}$ can be obtained from $\overline{C_{4m_{ij}+1}}$ by deleting $m_{ij}$ vertices, it follows by Property 2 that $S''_{ij}$ can be partitioned into 2 but no less than 2 cliques. Observe that the $k$ subgraphs $S_i$ and the $\binom{k}{2}$ subgraphs $S''_{ij}$ do not have edges in between. Since $k' = 3k + 2\binom{k}{2}$, we must partition each subgraph $S_i$ into exactly 3 cliques, and partition each subgraph $S''_{ij}$ into exactly 2 cliques. The remaining 2-intervals $\langle ue \rangle$ and $\langle ve \rangle$ for the edges $e$ are then added to these cliques. For each pair of distinct colors $i$ and $j$, $1 \le i < j \le k$, since $S'_{ij}$ is isomorphic to $\overline{C_{4m_{ij}+1}}$, it follows by Property 1 that there exists at least one edge $uv = e \in E_{ij}$ such that neither $\langle ue \rangle$ nor $\langle ve \rangle$ is included in the two cliques for $S''_{ij}$. Then $\langle ue \rangle$ must be included in one of the three cliques for $S_i$ that includes $\langle u \rangle$ (and $\langle ve \rangle$ must be included in one of the three cliques for $S_j$ that includes $\langle v \rangle$). Since $\langle ue \rangle$ intersects $\langle u \rangle$ but not the other 2-intervals in $S_i$, this clique includes only one 2-interval $\langle u \rangle$ from $S_i$. By Property 3, at most one of the three cliques for $S_i$ can include only one 2-interval from $S_i$. Now for each color $i$, $1 \le i \le k$, find the unique vertex $u_i$ such that the 2-interval $\langle u_i \rangle$ appears in a clique without any other 2-intervals from $S_i$. Then the set of $k$ vertices $u_i$ corresponds to a $k$-multicolored clique in $G$. □

## 9 Separating Vertices

In this section we prove Theorems 12, 13, 14, and 15. We use the notation $(a, b)$ to represent a 2-track interval where $a$ and $b$ are intervals on different tracks. We use similar notations for 3-track intervals.

**Proof of Theorem 12.** Following the approach of Marx [24], we show that $k$-SEPARATING VERTICES in balanced 2-track interval graphs is W[1]-hard with parameters $k$ and $l$ by an FPT reduction from $k$-CLIQUE.

Let $G = (V, E)$ be an input instance of $k$-CLIQUE with $n$ vertices and $m$ edges. We construct a family $\mathcal{F}$ of balanced 2-track intervals as shown in Figure 11, and an input instance $(G_\mathcal{F}, k', l')$ for $k$-SEPARATING VERTICES with $k' = k$ and $l' = 2\binom{k}{2}$.

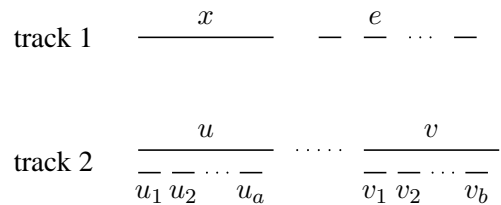

Figure 11: An illustration of the construction of $G_\mathcal{F}$ in Theorem 12.

On track 1 there are $m + 1$ disjoint intervals. The first interval, labeled by $x$, has length $n$; the other $m$ intervals, one for each edge $e \in E$, have length 1. On track 2 there are two rows of intervals. The first row has $n$ disjoint intervals of length $n$, one for each vertex in $V$. For a vertex $u \in V$, if the degree of $u$ is $a$, then there are $a$ disjoint intervals $u_1, u_2, \ldots, u_a$ of length 1 on the second row, all intersecting with the interval for $u$ in the first row.

There are $n + 2m$ balanced 2-track intervals in $\mathcal{F}$. For every vertex $u \in V$, add a 2-track interval $(x, u)$ to $\mathcal{F}$. For every vertex $u$, since there are $a = deg(u)$ many edges incident to $u$, fix an one-to-one correspondence between edges incident to $u$ and intervals $u_i$ with $1 \le i \le a$. For an edge $e = \{u, v\}$, let $u_i$ ($1 \leq i \leq a$) and $v_j$ ($1 \leq j \leq b$, where $b$ is the degree of $v$) be the intervals associated with $e$, add two 2-track intervals $(e, u_i)$ and $(e, v_j)$ to $\mathcal{F}$.

From the construction of $G_\mathcal{F}$, it is clear that $G_\mathcal{F}$ has a clique of size $n$, represented by the set of 2-track intervals $\{(x, u) \mid u \in V\}$. For an edge $e = \{u, v\}$ in $G$, $G_\mathcal{F}$ has a path of length three, represented by 2-track intervals $(x, u), (e, u_i), (e, v_j), (x, v)$, with the middle two vertices being degree-two.

If there is a $k$-clique $K$ in $G$, then we can cut the set of $k$ vertices in $G_\mathcal{F}$ represented by $\{(x, u) \mid u \in K\}$. By doing this, we separate $2\binom{k}{2}$ vertices represented by $\{(e, u), (e, v) \mid e \in E, e = \{u, v\}\}$. For the other direction, suppose $k'$ vertices can be deleted from $G_\mathcal{F}$ such that $l'$ vertices are separated from the rest of $G_\mathcal{F}$. We partition $k'$ deleted vertices into two parts $X$ and $Y$. Let $X$ be the set of vertices from the clique of size $n$ in $G_\mathcal{F}$ and $Y$ be the set of degree-two vertices in $G_\mathcal{F}$. Assume $n > k + 2\binom{k}{2}$, after deleting $X$ the rest of the clique in $G_\mathcal{F}$ has size greater than $l'$, so the $l'$ separated vertices must be degree-two vertices in $G_\mathcal{F}$. It is easy to see that by deleting $X$ at most $2\binom{|X|}{2}$ degree-two vertices are separated from the rest of $G_\mathcal{F}$, and by deleting $Y$ at most $|Y|$ degree-two vertices are separated from the rest of $G_\mathcal{F}$. Thus we have $|X| + |Y| = k$ and $2\binom{|X|}{2} + |Y| \geq 2\binom{k}{2}$. When $k \geq 2$, these conditions hold only when $|X| = k$ and $|Y| = 0$. This implies that the set of $k$ vertices $\{u \mid (x, u) \in X\}$ induces a clique in $G$.

**Proof of Theorem 13.** The reduction is also from $k$-CLIQUE. Given an input instance $G = (V, E)$ with $n$ vertices and $m$ edges for $k$-CLIQUE, we construct a family $\mathcal{F}$ of 3-track intervals as shown in Figure 12. We then show $k$-CLIQUE reduces to $k$-SEPARATING VERTICES in $\overline{G_\mathcal{F}}$.

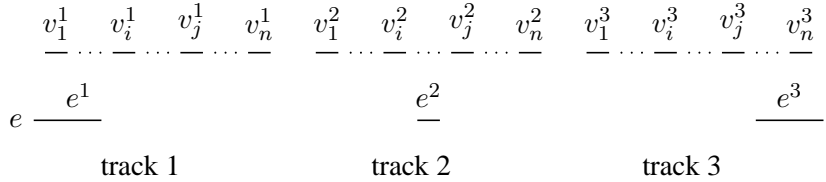

Figure 12: An illustration of the construction of $G_\mathcal{F}$ in Theorem 13. Only the 3-track interval $(e^1, e^2, e^3)$ corresponding to one edge $e$ is drawn.

Fix an ordering $v_1, \ldots, v_n$ of the vertices in $G$. On track $k$ ($1 \leq k \leq 3$), there are $n$ disjoint intervals $v_1^k, \ldots, v_n^k$. For every vertex $v_i \in V$, add a 3-track interval $(v_i^1, v_i^2, v_i^3)$ to $\mathcal{F}$. For every edge $e = \{v_i, v_j\}$ with $i < j$, add a 3-track interval $(e^1, e^2, e^3)$ (see Figure 12) to $\mathcal{F}$, such that $e^1$ intersects with $v_l^1$ for all $l < i$ on track 1, $e^2$ intersects with $v_l^2$ for all $i < l < j$ on track 2, and $e^3$ intersects with $v_l^3$ for all $l > j$ on track 3. The 3-track intervals for edges are pairwise intersecting at both left endpoint on track 1 and right endpoint on track 3.

It is clear that $\overline{G_\mathcal{F}}$ has a clique of size $n$, represented by the set of 3-track intervals $\{(v_i^1, v_i^2, v_i^3) \mid v_i \in G\}$. For each edge $e = \{v_i, v_j\}$ in $G$, $\overline{G_\mathcal{F}}$ has a path of length two, represented by 3-track intervals $(v_i^1, v_i^2, v_i^3), (e^1, e^2, e^3), (v_j^1, v_j^2, v_j^3)$ with the middle vertex $(e^1, e^2, e^3)$ being degree-two. Set $k' = k$ and $l' = \binom{k}{2}$. The rest of the proof is similar to the proof of Theorem 12.

For the sake of simple illustration, we did not draw the 3-intervals as balanced 3-intervals in Figure 12. Now we show how to transform them into balanced 3-intervals. First make all intervals of the form $v_i^k$ ($1 \leq i \leq n, 1 \leq k \leq 3$) unit-length open intervals. On track 1 align them next to each other without any gap between $v_i^1$ and $v_{i+1}^1$ for all $1 \leq i < n$. Do the same for track 3. But on track 2, align them with a gap of length $n$ between $v_i^2$ and $v_{i+1}^2$ for all $1 \leq i < n$. Then, for any edge $e = \{v_i, v_j\}$ with $i < j$, we can always use a balanced 3-track interval $(e^1, e^2, e^3)$ to achieve the same intersecting pattern as shown in Figure 12. In particular, first choose an appropriate length (between $n$ and $n^2$) for $e^2$ so that $e^2$ intersects with $v_l^2$ for all $i < l < j$ on track 2, then make $e^1$ and $e^3$ the same length by extending $e^1$ to the left and $e^3$ to the right if necessary.

**Proof of Theorem 14.** For the W[1]-hardness in balanced 2-track interval graphs, we use the same construction as in the proof of Theorem 12, and ask whether $l = n + 2m - 2\binom{k}{2} - k$ connected vertices can be separated from $G_\mathcal{F}$ by deleting $k$ vertices. Similarly, for the W[1]-hardness in co-balanced 3-track interval graphs, we use the same construction as in the proof of Theorem 13, and ask whether $l = n + m - \binom{k}{2} - k$ connected vertices can be separated from $\overline{G_\mathcal{F}}$ by deleting $k$ vertices.

**Proof of Theorem 15.** Use the same constructions as in the proofs of Theorem 12 and Theorem 13. Ask whether $G_\mathcal{F}$ (or $\overline{G_\mathcal{F}}$) can be separated into $l = \binom{k}{2} + 1$ components by deleting $k$ vertices.

## 10 Irredundant Set

In this section we prove Theorem 16. We show that $k$-IRREDUNDANT SET is W[1]-hard by an FPT reduction from the W[1]-complete problem $k$-MULTICOLORED CLIQUE [10].

Let $(G, \kappa)$ be an instance of $k$-MULTICOLORED CLIQUE. We will construct a graph $G'$ such that $G$ has a clique of $k$ vertices containing exactly one vertex of each color if and only if $\overline{G'}$ has an irredundant set of $k'$ vertices, where $k' = 3k + 5\binom{k}{2}$.

*Vertex Selection*: For each color $i$, $1 \le i \le k$, the graph $G'$ contains a subgraph $G'_i$ as the *vertex gadget* for the color $i$. Let $V_i$ be the set of vertices in $G$ with color $i$. For each vertex $u \in V_i$, $G'_i$ includes 3 vertices $u_1, u_2, u_3$ forming a 3-clique. The vertices from different 3-cliques in $G'_i$ are disjoint.

*Edge Selection*: For each pair of distinct colors $i$ and $j$, $1 \le i < j \le k$, the graph $G'$ contains a subgraph $G'_{ij}$ as the *edge gadget* for the color pair $ij$. Let $E_{ij}$ be the set of edges $uv$ such that $u$ has color $i$ and $v$ has color $j$. For each edge $e = uv \in E_{ij}$, $G'_i$ includes 5 vertices $e_1, e_2, e_3, e_4, e_5$ forming a 5-clique. The vertices from different 5-cliques in $G'_{ij}$ are disjoint.

*Validation*: Each edge gadget $G'_{ij}$ is connected to the two vertex gadgets $G'_i$ and $G'_j$ as follows. For each edge $e = uv \in E_{ij}$, each of the 5 vertices $e_1, e_2, e_3, e_4, e_5$ is connected to each of the 3 vertices $u_1, u_2, u_3$ and to each of the 3 vertices $v_1, v_2, v_3$. In addition, we connect the edge gadget $G'_{ij}$ to each vertex gadget $G'_z$, $z \in \{1, 2, \ldots, k\} - \{i, j\}$, by adding all possible edges between them. Also, we connect different edge gadgets to each other, and connect different vertex gadgets to each other, by adding all possible edges between them.

**Lemma 7.** *$G$ has a clique of $k$ vertices containing exactly one vertex of each color if and only if $\overline{G'}$ has an irredundant set of $k'$ vertices, where $k' = 3k + 5\binom{k}{2}$.*

*Proof.* We first prove the direct implication. Suppose that $G$ has a clique $K$ of $k$ vertices containing exactly one vertex of each color. Let $I$ be the set of $k'$ vertices in $G'$ including the 3 vertices $u_1, u_2, u_3$ for each vertex $u \in V(K)$ and the 5 vertices $e_1, e_2, e_3, e_4, e_5$ for each edge $e \in E(K)$. Observe that $I$ is a clique in $G'$. It follows that $I$ is an independent set hence also an irredundant set in $\overline{G'}$.

We next prove the reverse implication. Suppose that $\overline{G'}$ has an irredundant set $I$ of $k'$ vertices. We start with two simple propositions:

1. For each color $i$, $I$ includes at most 3 vertices in the subgraph $\overline{G'_i}$. Moreover, if $I$ includes exactly 3 vertices in $\overline{G'_i}$, then they must be the vertices $u_1, u_2, u_3$ from a 3-clique in $G'_i$ corresponding to some vertex $u \in V_i$.

2. For each color pair $ij$, $I$ includes at most 5 vertices in the subgraph $\overline{G'_{ij}}$. Moreover, if $I$ includes exactly 5 vertices in $\overline{G'_{ij}}$, then they must be the vertices $e_1, e_2, e_3, e_4, e_5$ from a 5-clique in $G'_{ij}$ corresponding to some edge $e \in E_{ij}$.

To prove the first proposition, observe that any two vertices in the same 3-clique in $G'_i$ has the same open neighborhood in $\overline{G'}$. If $I$ includes two or more vertices from the same 3-clique in $G'_i$, then all these vertices must be self-private, and $I$ cannot include any vertex from a different 3-clique in $G'_i$. Suppose that $I$ includes three or more vertices in $G'_i$ that are not all from the same 3-clique, then these vertices must come from distinct 3-cliques in $G'_i$. Let $\alpha, \beta, \gamma$ be three such vertices. Observe that $\gamma$ is adjacent to both $\alpha$ and $\beta$ in $\overline{G'_i}$. Also observe that the open neighborhood of $\gamma$ in $\overline{G'}$ is contained in the union of the open neighborhoods of $\alpha$ and $\beta$ in $\overline{G'}$. Thus $\gamma$ cannot have a private neighbor, self-private or not. Similarly for $\alpha$ and $\beta$. This contradicts their membership in $I$.

To prove the second proposition, observe that any two vertices in the same 5-clique in $G'_{ij}$ has the same open neighborhood in $\overline{G'}$. If $I$ includes two or more vertices from the same 5-clique in $G'_{ij}$, then all these vertices must be self-private, and $I$ cannot include any vertex from a different 5-clique in $G'_{ij}$. Suppose that $I$ includes five or more vertices in $G'_{ij}$ that are not all from the same 5-clique, then these vertices must come from distinct 5-cliques in $G'_{ij}$. Let $\alpha, \beta, \gamma, \mu, \nu$ be five such vertices. These vertices are pairwise adjacent in $\overline{G'_{ij}}$, so they cannot be self-private. Observe that within the subgraph $\overline{G'_{ij}}$, the open neighborhood of each of these five vertices is contained in the union of the open neighborhoods of any two of the other four vertices. Also observe that within any gadget subgraph except $\overline{G'_i}$, $\overline{G'_j}$, and $\overline{G'_{ij}}$, any two of these five vertices have the same (empty) open neighborhood. From these observations, it follows that these five vertices must have private neighbors in $\overline{G'_i}$ and $\overline{G'_j}$. Then, at least three of the five vertices must have private neighbors either all in $\overline{G'_i}$ or all in $\overline{G'_j}$. Assume without loss of generality that the three vertices $\alpha, \beta, \gamma$ have private neighbors in $\overline{G'_j}$. If any two of the three vertices have the same open neighborhood in $\overline{G'_i}$, then the two vertices cannot both have private neighbors in $\overline{G'_i}$. Otherwise, the open neighborhood of any one of the three vertices is contained in the union of the open neighborhoods of the other two, so none of the three vertices can have a private neighbor in $\overline{G'_i}$. We have reached a contradiction.

There are exactly $k$ vertex gadgets and exactly $\binom{k}{2}$ edge gadgets in our construction. Note that $k' = 3k + 5\binom{k}{2}$. From the two propositions, it follows that $I$ must include exactly 3 vertices $u_1, u_2, u_3$ from each vertex gadget $G_i$ corresponding to a vertex $u \in V_i$, and exactly 5 vertices $e_1, e_2, e_3, e_4, e_5$ from each edge gadget $G_{ij}$ corresponding to an edge $e \in E_{ij}$. Moreover, these $k'$ vertices are all self-private, so the irredundant set $I$ is indeed an independent set in $\overline{G'}$. Then the corresponding $k$ vertices and $\binom{k}{2}$ edges in $G$ must be consistent, forming a multicolored clique with exactly one vertex of each color. □

## 11 Concluding Remarks

Although we have managed to devise a simpler proof for the W[1]-hardness of $k$-IRREDUNDANT SET in general graphs, we were unable to strengthen this result by proving the W[1]-hardness of $k$-IRREDUNDANT SET in $t$-interval graphs or co-$t$-interval graphs for any constant $t$. Both the graph in the previous proof of Downey et al. [10] and the graph in our simpler proof contain very large complete bipartite graphs and complements of complete bipartite graphs. It is known [17] that the interval number of the complete bipartite graph $K_{\lfloor n/2 \rfloor, \lceil n/2 \rceil}$ is $\lceil \frac{n+1}{4} \rceil$, i.e., $\lceil \frac{n+1}{4} \rceil$ is the smallest number $t$ such that $K_{\lfloor n/2 \rfloor, \lceil n/2 \rceil}$ is a $t$-interval graph. Therefore, unless with new techniques, the existing constructions cannot be directly adapted to prove the W[1]-hardness of $k$-IRREDUNDANT SET in $t$-interval graphs or co-$t$-interval graphs even if $t$ is a parameter of the problem besides $k$.

A general direction for extending our work is to strengthen the existing hardness results for more restricted graph classes. For example, we showed in Theorem 2 that $k$-DOMINATING SET in co-3-track interval graphs is W[1]-hard with parameter $k$. Is it still W[1]-hard in co-2-track interval graphs or co-unit 3-track interval graphs? Many questions can be asked in the same spirit. In particular, are $k$-INDEPENDENT DOMINATING SET and $k$-PERFECT CODE W[1]-hard in co-$t$-interval graphs for some constant $t \geq 2$?